\documentclass[reqno,12pt]{iopart}
   \usepackage{iopams}

\begin{document}

\title[Unifying the Construction of Various Types of Generalized
 Coherent States]{PhD Tutorial: Unifying the Construction of Various Types of Generalized
 Coherent States}

\author{M. K. Tavassoly }
 \eads{\mailto{mk.tavassoly@yazduni.ac.ir}}
 \small{ 1. Quantum Optics Group, Department of Physics,
  University of Isfahan, Isfahan, Iran  }\\
\small{2. Department of Physics, University of Yazd, Yazd, Iran}

\newcommand{\I}{\mathbb{I}}

\newcommand{\norm}[1]{\left\Vert#1\right\Vert}
\newcommand{\abs}[1]{\left\vert#1\right\vert}
\newcommand{\set}[1]{\left\{#1\right\}}
\newcommand{\R}{\mathbb R}
\newcommand{\C}{\mathbb C}
\newcommand{\eps}{\varepsilon}
\newcommand{\To}{\longrightarrow}
\newcommand{\BX}{\mathbf{B}(X)}
\newcommand{\HH}{\mathfrak{H}}
\newcommand{\D}{\mathcal{D}}
\newcommand{\N}{\mathcal{N}}
\newcommand{\la}{\lambda}
\newcommand{\af}{a^{ }_F}
\newcommand{\afd}{a^\dag_F}
\newcommand{\afy}{a^{ }_{F^{-1}}}
\newcommand{\afdy}{a^\dag_{F^{-1}}}
\newcommand{\fn}{\phi^{ }_n}
 \newcommand{\HD}{\hat{\mathcal{H}}}
\maketitle

\begin{abstract}
In this tutorial I intend to present some of the results I
obtained through my PhD work in the "quantum optics group of the
University of Isfahan" under consideration Dr. R. Roknizadeh and
Prof. S. Twareque Ali as my supervisor and advisor, respectively.
I will revisit some of the pioneering proposals recently
developed the concept of generalized CSs. As it can be observed
the customary three generalization methods {\it (symmetry,
algebraic and dynamical)}  have never been considered in neither
of them. Our intention in this work is at first to investigate
the lost ring between the customary three methods and the recently
developed ones, as possible.  For this purpose it has been
devised general analytic descriptions, which successfully
demonstrate how different varieties of CSs (which are nonlinear
in nature) can be obtained by two processes, first the {\it
"nonlinear CSs"} method and second by {\it "basis transformations
on an underlying Hilbert spaces"}.   As a result, I will
systematize the recently introduced generalized CSs in a clear
and concise way. It will be clear also, that how our results can
be considered as a first step in the generation process of the
mathematical physics CSs in the context of quantum optics.
Besides this, some new results emerge from our studies. I
introduce a large classes of generalized CSs, namely the {\it
"dual family"} associated with each set of early known CSs. But,
in this relation, the previous processes for constructing the
dual pair of Gazeau-Klauder CSs fail to work well, so I outlined
a rather different method based on the {\it "temporal stability"}
requirement of generalized nonlinear CSs.

\end{abstract}
{\bf keyword:} {nonlinear coherent states, generalized coherent
     states}
\submitto{\JOB}
\section{Introduction}\label{sec-intro}
 Quantum optics is an ideal testing ground for ideas of quantum
 theory, and coherent states (CSs),
  first introduced by Shr\"{o}dinger \cite{Shrodinger} as  venerable objects in
  physics, are one of the basic axes in researches in this branch of
 physics. Nowadays much
 attention is being paid to the CSs and their theoretical
 generalizations \cite{Alia, AAG-book, klauskag, Odzijewicz}, including their experimental generations and
 applications. The new generalizations have some interesting
 {\it "non-classical
 properties"} such as squeezing, antibunching, sub-Poissonian
 statistics and oscillatory number distribution. Earlier study of
 such non-classical effects were regarded as being of interest
 academically only, but now their applications in quantum communications
 \cite {Schumachera, Schumacherb}, quantum teleporation
 \cite {Braunstein_Ariano, Bennett_1, Braunstein_Kimble}, dense coding \cite {Braunstein},
 quantum cryptography \cite {Bennett_1, Kemmpe} and detection of gravitational
 waves \cite{Giacoboin, Kimble}
 are well understood.

 It is frequently found in the literature that motivations to
 generalize the concept  of CSs have arisen from {\it "three
 methods", symmetry considerations, dynamics and  algebraic aspects}.
 Generalization based on symmetry has led to define CSs for
 arbitrary Lie groups \cite {Brifa, Brifb, Brifc, Perelomov1, Perelomov2}.
 Based on dynamics, CSs have been
 constructed for systems other than harmonic oscillator which has equally spaced energy
 levels \cite {Nieto-a, Nieto-b, Nieto-c, Nieto-Sim78} and finally
 CSs for deformed algebras have been introduced by
 extending the algebraic definition \cite{Arik, Manko1997, Shanta5}.

 But it is also well-known that  there are generalizations in very many ways
 that (in the first sight) have not derived by the above three
 approaches. To say a few, we can refer to even and odd CSs \cite{Dodonov},
 Shr\"{o}dinger cat
 states \cite{Shrodinger2, Yurke}, all of which have been obtained
  from some special {\it superpositions} of
 canonical CSs or other generalizations.
    While the canonical CSs do not have any non-classical properties, their
    superpositions have.
    Another scheme for generalization of CSs are obtained
    by using the non-orthogonal bases of a Hilbert space.
    Part of our works contains the construction of CSs
    and squeezed state(SSs) using a special set of non-orthogonal but normalizable bases
    $\{|n\rangle_\lambda, n \in \mathbf{N}\}_{n=0}^\infty$ instead of
    orthonormal one $\{|n\rangle, n \in \mathbf{N}\}_{n=0}^\infty$ \cite{rokntvs}.

 In addition to the above generalizations, recently there are
 some new classes of generalizations of CSs, such as the
 ones introduced  by Klauder-Penson-Sixdeniers (KPS) \cite{kps},
 generalized hypergeometric (HG) CSs \cite{Appschill},
 Gazeau-Klauder CSs (GK-CSs)
  \cite{Gazeau-Monceau2000, gazklau, Klauder96, Klauder98, Klauder2001},
 Penson-Solomon (PS) CSs \cite{Penson}, including the
 Mittag-Leffler (ML) CSs \cite{Mittag} and Tricomi (TC) CSs \cite{spk}
 have been constructed by different {\it "mathematical structures"}.
 In neither of these works the authors referred to any of the {\it "three customary methods of generalizations"}.
 In some of them, even the physical motivations of the three generalizations
 have been questioned essentially \cite{Klauder2001}.

  As is well known,  the idea of {\it "unifying"} the different
  methods in any field of physics is an old subject which is of
  interest,  as a major challenge for theoretical physicists.
  The basic aim in this manufacture is to present formalisms to
  {\it "unify the different ways of the above scattered models
  of generalizations of CSs"} \cite{tavassoly}.
  Through doing this, we will arrive at some new families of generalized CSs
  in the context of quantum optics.
    In addition to the {\it elegance} of the present work, this classifications will simplify
  understanding and the introduction of the  theories,  more than before.
  To achieve this purpose,  two different approaches will be presented.
  \begin{enumerate}
 \item {\it The nonlinear CSs formalism,}\\
   The first approach,   basically based on the
   conjecture that all these states may be
   studied  in the so-called {\it "nonlinear CSs"}
   or {\it "$f$-deformed CSs"} category \cite {Manko1997, Matos1996,
   MMatos1996}, the states that
   attracted much attention in recent years, mostly because they exhibit
   nonclassical properties. Up to now, many quantum
   optical states such as $q$-deformed CSs \cite{Manko1997}, negative
   binomial state \cite{Wang-Fu, Wang-Comm}, photon added (and subtracted)
   CSs \cite{Siv1999, Siv2000, Naderi, Naderi2004b, Naderi-pacs}, the center of mass motion
   of a trapped ion \cite{Matos1996, MMatos1996}, some nonlinear phenomena such as
   a hypothetical {\it "frequency blue shift"} in high intensity
   photon beams \cite{MankoTino} and recently after proposing
   $f$-bounded CSs \cite{Recamier}, the binomial state (or
   displaced exited CSs) \cite{rokntvs} have been considered as
   some sorts of nonlinear CSs.
   Speaking otherwise, the algebraic
   method may be particularly useful for providing a {\it unified} treatment of
   all these states and their {\it interrelationships}.
   I attempt now to demonstrate that all sets of KPS and ML CSs, HG,
   PS and TC CSs, including  the two discrete
   series representations of the group $SU(1, 1)$
   (both Barut-Girardello (BG) \cite{bar-gir71} and Gilmore-Perelomov (GP) CSs \cite{gilmore}) and $su(1, 1)$-BG CSs for
   Landau-level (LL) \cite{Fakhri} can be classified in the nonlinear CSs with some
   special nonlinearity function, $f(n)$, by which we may obtain the {\it "deformed
   annihilation and creation operators"}, {\it "generalized displacement
   operator"} and the {\it "dynamical Hamiltonian"}  corresponding to each class of them.
   Based on these results, it will be possible to reconstruct
   all of the above CSs via conventional
   fashions, i.e. by annihilation and displacement (type) operator
   definition.

   \item
 {\it The mathematical physics formalism,}\\
 In this approach we attempt to
 set up a mathematical formalism by which one can produce a vast
 class of generalized CSs, e.g.
 binomial states, photon-added CSs and the so-called nonlinear CSs. Therefore, all of the generalized CSs
 that by the first approach we clarified the nonlinearity
 nature's, can also be reconstructed by the second approach  \cite{AliRokTav}.
 By this formalism we try to establish that a vast class of generalized
 CSs can be considered as different representations of canonical CSs in
 the underlying  Hilbert spaces.
  Indeed by
 defining an appropriate operator $\hat T$ (with some special properties)
 and its action on the canonical CSs one may yield the above generalized CSs.
 As we will observe later, for the case of nonlinear CSs this
 operator depends explicitly on the nonlinearity function $f(\hat n)$.
 Therefore, in a sense it enables one to obtain some various
 types of generalized CSs via a {\it single} mathematical formalism, rather than various distinguishable approaches.
 Obviously, in each case we have to find an appropriate $\hat T$ operator.
\end{enumerate}

   In another direction, as a matter of fact
   introducing the deformed creation and annihilation operators related to the mathematical physics CSs
   considered in this manufacture may be regarded as a first step in the process of production and detection of
   these states in the experimental realization schemes of quantum
   optics.
   According to the  theoretical scheme proposed in  Ref. \cite{Naderi2004a}, any
   nonlinear CSs can be generated in a micromaser under
   intensity-dependent  (ID) Jaynes-Cummings model.
   The authors generalized the "standard"  Jaynes-Cummings
   model to the "multi-photon intensity-dependent" case, which may be
   defined as a quantum model describing the interaction of a
   monochromatic electromagnetic field with one two-level atom in
   a cavity under intensity-dependent coupling through
   multi-photon transitions. The interaction Hamiltonian of this
   model can be expressed in the rotating-wave approximation and in the interaction picture as follows ($\hbar=1$):
  \begin {equation} \label{ID-m}
  \hat H ^{(m)}_{ID} = g \left[ a^m f(\hat n) |a\rangle \langle b |
  + |b\rangle \langle a| f(\hat n) (a^\dagger)^m \right], \qquad
  m=1,2, \cdots \;,
  \end {equation}
  where  $|a\rangle$ and $|b\rangle$    denote the exited and ground states of atomic
  level,respectively,
  $g$ is the coupling constant, $a$, $a^\dagger$ are the standard photon annihilation and
  creation  operators with algebra $[a, a^\dagger]=\hat I$ and $f(\hat n)$ describes the intensity dependence of atom-field interaction.
  As is well known $f(\hat n)$ is the same operator valued (or {\it "nonlinearity"}) function
  associated with any class of nonlinear CSs. Although it is
  pointed in \cite{Naderi2004a} that $f(\hat n)$ is a real
  function, in my opinion the formalism can also be extended to any phase-dependent one (the case that I will strict in
  my researches in the Gazeau-Klauder generalized CSs).

  For the special case of one-photon transition the model equation
  (\ref {ID-m}) simplify to
  \begin {equation} \label{ID-1}
     \hat H ^{(m=1)}_{ID} =g \left[ A |a\rangle \langle b |
    + |b\rangle \langle a| A^\dagger \right], \qquad
     m=1,2, \cdots \;,
    \end {equation}
  where in the latter equation we have set $A=af(\hat n)$ and
  $A^\dagger = a^\dagger f(\hat n)$, which are the known $f-$deformed
  ladder operators as the generators of the nonlinear oscillator algebras $\{A, A^\dagger, \hat H\}$.
  {\it Therefore, as we will observe, the results obtained from the first approach can be considered as
  an introductory step in the theoretical scheme for "generation of the
  mathematical physics CSs" have been developed in recent decade.}

   In addition to these, some interesting and new remarkable points
   emerge from  the presented results and studies.
   The first is that both of the two mentioned formalisms provide a framework to construct a vast new families of
   CSs (named the {\it dual family}), other than   KPS, GH, ML, PS and
   TC CSs.
   The second is that the
   Hamiltonian proposed in \cite{Manko1997} and others who cited to
   him (see e.g. \cite{Siv2000}) must be reformed, in view of the {\it
   action identity} requirement imposed on the generalized CSs
   proposed by Gazeau and Klauder \cite{gazklau, Klauder2001}. We should quote here that recently some
   authors (for instance see
   \cite{ElKinani2001b, ElKinani2002, ElKinani20022, ElKinaniand_Daoud, Elkinani2003})
   have used normal ordering form (factorization) for their
   Hamiltonians. Indeed they used the {\it "supersymmetric quantum mechanics"} (SUSQM)
   techniques \cite{Witten} as a {\it "mathematical tool"} to find the ladder operators for their Hamiltonians.
   Interestingly our formalism for solvable Hamiltonians gives an
   easier and clearer manner to obtain these operators whenever  necessary.

 As we will observe, the above two approaches are not enough to reproduce all the known generalized CSs.
 GK-CSs is an important case \cite{gazklau}.
 Applying the first formalism to the GK-CSs, gives readily us a function $f(\hat n)$
 and therefore by the second one we will find an operator $\hat T(n)$. But both of these two operators are
 ill defined, because they  depend on a  variable $\gamma$  ($-\infty < \gamma < \infty$).
 Other than this defect, we will observe that the dual family of
 GK-CSs obtained by each of the two approaches are not fully
 consistent with the Gazeau-Klauder criteria. Strictly speaking
 it is easily found that they did not satisfy the temporal stability (and therefore the action
 identity) requirements.
 To overcome this serious problem, upon using the analytical representations of GK-CSs (denoted by GKCSs),
 I will introduce
 an operator $\hat S$, depends explicitly on the Hamiltonian of the system,
 which its action on any nonlinear CSs (which essentially do not possess the temporal stability property), transfers
 it to a situation that enjoy this property, nicely
 \cite{Roknizadeh-Tav-AIP}.
 In this way, not only we are able to derive the dual family of GKCSs, but
 also we may obtain the opportunity to introduce the temporal
 stability version of all the nonlinear CSs, derived in this
 tutorial  and introduced elsewhere.

   This tutorial organizes as follows: in section 2 we revisit
   some of the most important generalized CSs of recent decade,
   which will be as a  necessary tool for our works in later sections.
   Along unification of mathematical physics (MP) CSs introduced in section 2, section 3 deals with the two unification
   methods successfully reach this purpose.
   The generalized displacement operator and the dual family of
   some of the MP CSs will be presented in section 4.
   The dual family of GKCSs will be
   presented in section 5 in a general framework of the
   temporal stabilization of nonlinear CSs, with some physical appearance of the presented formalism.
   Finally in section 6 a scheme for the generalized GKCSs and the
   associated dual family will be offered.

\section{Preliminary arguments on mathematical physics (MP)
 CSs }
 In this section we will express the explicit form of some generalized CSs
 which are necessary in our future treatments.
 Nearly all of these  generalized CSs (except the GK-CSs, GKCSs and $su(1, 1)$-BG CSs for Landau levels),
 have been introduced based on some {\it "mathematical
 structures"}. So we have called whole of them as {\it "mathematical physics
 CSs"}.

\subsection{ Nonlinear CSs and their dual family}\label{sec-nl}
    Nonlinear CSs first introduced explicitly in \cite{filvog, Manko1997,  Matos1996,
    MMatos1996},  but before them it is implicitly defined by
    Shanta {\it etal} \cite{Shanta5} in a compact form.
    This notion  attracted much attention in physical literature  in
    recent decade, especially because of their nonclassical
    properties in quantum optics. Man'ko {\it etal}'s approach is based on the two
    following postulates.

    The first is that the standard annihilation and
    creation operators deformed with an intensity dependent function
    $f(\hat{n})$ (which is an operator valued function), according to
    the relations:
  \begin{equation}\label{nonl-annih}
     A=af(\hat{n})=f(\hat{n}+1)a, \qquad A^\dag=f^{\dag}(\hat{n})a^\dag = a^{\dag}
     f^{\dag}(\hat{n}+1),
  \end{equation}
       with commutators between $A$ and $A^\dag$ as
  \begin{equation}\label{comut}
    [A,A^\dag]=(\hat{n}+1)f(\hat{n}+1)f^\dag(\hat{n}+1)-\hat{n}f^\dag(\hat{n})f(\hat{n}),
  \end{equation}
    where $a$, $a^\dag$ and
    $\hat{n}=a^\dag  a$ are bosonic annihilation, creation and number
    operators, respectively. Ordinarily the phase of $f$ is irrelevant and
    one may choose $f$ to be real and nonnegative, i.e.
    $f^\dag(\hat{n})=f(\hat{n})$. But to keep general
    consideration, we take into account the phase dependence of
    $f(\hat{n})$ in general formalism given here.

    The second postulate is that the Hamiltonian of the deformed oscillator in analog to the
    harmonic oscillator is found to be
 \begin{equation}\label{hamilt}
   \hat{H}_{\rm M}=\frac{1}{2}(A A^\dag+A^\dag A),
 \end{equation}
   which by Eqs. (\ref{nonl-annih})  can be
   rewritten as
 \begin{equation}\label{hamilt-f}
   \hat{H}_{\rm M}=\frac{1}{2}\left((\hat{n}+1)f(\hat{n}+1)f^\dag(\hat{n}+1)+\hat{n}
   f^\dag(\hat{n})f(\hat{n})\right),
 \end{equation}
   where index $M$ refers to the Hamiltonian as introduced by Man'ko \etal \cite{Manko1997}.

   The single mode nonlinear CSs obtained as eigen-state of the
   annihilation operator is as follows:
 \begin{equation}\label{nonl-cs}
   |z, f\rangle = \N_f(|z|^2)^{-1/2}\sum_{n=0}^{\infty}C_n z^n
   |n\rangle,
 \end{equation}
   where the coefficients $C_n$ are given by
 \begin{equation}\label{Cn}
   C_n=\left(\sqrt{[nf^\dag(n)f(n)]!}\right)^{-1}, \quad C_0=1, \quad
   [f(n)]!\doteq f(n)f(n-1)\cdots f(1),
 \end{equation}
   and the normalization constant is determined as
   $ \N_f(|z|^2)=\sum_{n=0}^{\infty} {|C_n|^2 |z|^{2n}}.$
   In order to have states belonging to the Fock space, it is
   required that $0< \N_f(|z|^2) < \infty$, which implies
   $|z|\leq \lim_{n \mapsto \infty}n[f(n)]^2$.
   No further restrictions are then put on $f(n)$. Now with the
   help of Eqs. (\ref {nonl-cs}) and (\ref{Cn}) the function $f(n)$
   corresponding to any nonlinear CSs is found to be
 \begin{equation}\label{findf}
   f(n)=\frac{C_{n-1}}{\sqrt n C_n}\;,
 \end{equation}
   which plays the key rule in our present work.
   To recognize the nonlinearity of any CSs we can use this simple and
   useful relation; by this we mean that if $C_n$'s for any CSs are
   known, then $f(n)$ can be found from Eq. (\ref {findf}); when $f(n)=1$
   or at most be only a constant phase,
   we recover the original oscillator algebra, otherwise it is
   nonlinear.

  Recall that by replacing $f(n)$ with $\frac{1}{f(n)}$ in the
  relations (\ref {nonl-cs}) and (\ref {Cn}) one
  immediately gets the nonlinear CSs  introduced in \cite{royroy}.
  These latter states have been called as the {\it "dual family"} of nonlinear CSs of
  Man'ko's type \cite{AliRokTav}.

 \subsection{ Klauder-Penson-Sixdeniers (KPS) and Mittag-Leffler (ML) generalized CSs}\label{kps}
   Along generalizations of CSs,   Klauder, Penson and Sixdeniers  introduced the
   states \cite{kps}
  \begin{equation}\label{kps-cs}
    {|z\rangle_{\rm {KPS}}}=\N(|z|^2)^{-1/2}\sum_{n=0}^{\infty}\frac{z^n}
   {\sqrt{\rho(n)}}|n\rangle,
  \end{equation}
      where  $\rho(n)$ satisfies $\rho(0)=1$ and the normalization constant is determined as
     $\N(|z|^2)=\sum_{n=0}^{\infty}\frac{|z|^{2n}}{\rho(n)}.$
  These states possess three main properties:  (i) {\it normalization} (ii) {\it continuity in the  label}
   and  (iii)  {\it they form an over-complete set}
   which allows a resolution of the identity with a positive weight function,  by
  appropriately selected functions, $\rho(n)$. The third condition,
  which is the most difficult and at the same time the strongest
  requirements of any sets of CSs, were proved appreciatively by
  them, through Stieltjes and Hausdorff power moment
  problem. Explicitly for each particular set of generalized CSs ${|z\rangle_{_{KPS}}}$, they
  found the positive weight function $W(|z|^2)$ such that
 \begin{equation}\label{kps-res1}
    \int\int_{\C} d^2z {|z\rangle_{\rm {KPS}}} W(|z|^2)_{\rm {KPS}}\langle
    z|=\hat{I}=\sum_{n=0}^{\infty}|n\rangle \langle n|,
 \end{equation}
   where $d^2 z=|z|d|z|d\theta$. Strictly speaking, evaluating the integral in (\ref{kps-res1}) over
   $\theta$ in the LHS of Eq. (\ref {kps-res1}), setting $|z|^2\equiv
   x$ and simplify it, we arrive finally at:
 \begin{equation}\label{kps-res2}
   \int_0^R x^n W'(x)dx=\rho(n),  \qquad n=0, 1, 2, ...,   \qquad
   0<R\leq\infty,
 \end{equation}
    where the positive weight functions $W'(x)=\frac{\pi
    W(x)}{\N(x)}$ must be determined.
   Then, the authors used the general formalism to a variety of
   the $\rho(n)$ functions have been defined  on the {\it "whole plane"} and on
   the {\it "unit disk"}. In all cases the functions $\rho(n)$ have been
   choosed such that the resolution of the identity hold.

   A special case of these states known as ML CSs  may be constructed by replacing the
   function $\rho(n)=\frac{\Gamma(\alpha n
   +\beta)}{\Gamma(\beta)}$ in (\ref{kps-cs}), where $\alpha\;, \beta > 0$ \cite{Mittag}.

\subsection{ Generalized hypergeometric (HG) CSs}

   A larger class of KPS CSs, can be constructed
  by starting with the hypergeometric functions \cite{Appschill},
\begin{equation}\label{hypergeomfcn}
 _pF_q (\alpha_1, \ldots , \alpha_p ; \; \beta_1, \ldots , \beta_q ; \; x )
       = \sum_{n=0}^\infty \frac {(\alpha_1)_n \ldots (\alpha_p)_n}
           {(\beta_1)_n  \ldots (\beta_q)_n}\; \frac {x^n}{n!}\; ,
\end{equation}
where  $\alpha_i$ and $\beta_i$ are positive real numbers, $q$ is
an arbitrary positive integer and $p$ is restricted by $q-1 \leq
p \leq q +1$. (Here $(\gamma)_n$ is the usual Pochhammer symbol,
$(\gamma)_n = \gamma (\gamma+1)(\gamma+2)\ldots (\gamma+n-1) =
\Gamma (\gamma + n)/\Gamma (\gamma)$). The series in (\ref
{hypergeomfcn}) converges for all $x \in \R$ if $p = q$ and for
all $\vert x \vert < 1$ if $p = q+1$. The explicit form of these
states are as follows \cite{Appschill}:
\begin{equation}\label{GHCS}
| z; p, q \rangle _{\rm HG}= | \alpha_1, \cdots \alpha_p; \beta_1,
 \cdots \beta_q; z \rangle = {}_p {\N}_q (|z|^2) ^{-1/2}
\sum_{n=0}^\infty \frac{z^n}{\sqrt {_p \rho _q(n)}}\;|n\rangle,
\end{equation}
where $_p \rho _q(n)$ are the strictly positive functions of $n$,
defined by the relation
\begin{equation}\label{p-rho-q}
_p \rho _q(n) = {}_p\rho _q(\alpha_1, \cdots \alpha_p; \beta_1,
 \cdots \beta_q; n) = \Gamma(n+1)\frac{(\beta_1)_n \cdots
(\beta_q)_n}{(\alpha_1)_n \cdots (\alpha_p)_n},
\end{equation}
and the normalization factor is determined as
           ${}_p {\N}_q (|z|^2) = {}_pF_q (\alpha_1 , \ldots ,
            \alpha_p ; \; \beta_1 , \ldots , \beta_q ; \; |z|^2 ).$

\subsection{ Generalized Tricomi (TC) CSs}
Tricomi (TC) CSs of the first kind  introduced as \cite{spk}
\begin{equation}\label{TC-1}
\fl |z; p\rangle _{\rm TC}^{(1)}= \N_p(|z|^2)^{-1/2}
\sum_{n=0}^\infty \frac{z^n}{\sqrt{n! d_p(n)}}|n\rangle, \quad
d_p(n) \equiv \frac{p^{-\frac {n} {2}} 2^{-n}}{\sqrt \pi
e^{\frac{1}{4p}}erfc(\frac{1}{2\sqrt p})}\psi\left(
\frac{n+1}{2}\;, \frac 1 2;  \frac{1}{4p}\right),
\end{equation}
and similarly the second kind of Tricomi's CSs defined as
\cite{spk}
\begin{equation}\label{TC-2}
\fl |z; \lambda, \beta \rangle _{\rm TC}^{(2)}= \N_{\lambda,
\beta}(|z|^2)^{-1/2} \sum_{n=0}^\infty \frac{z^n}{\sqrt{n!
d_{\lambda, \beta}(n)}}|n\rangle, \quad d_{\lambda, \beta}(n)
\equiv \frac{\beta ^n \psi (n+1, n+2-\lambda; \beta)}{ \psi (1,
2-\lambda; \beta)},
\end{equation}
where $p, \beta > 0$, $\lambda$ is arbitrary, $\psi (a,c; z)$ are
the Tricomi's confluent hypergeometric functions and  the
normalization constants $\N_p$ and $\N_{\lambda, \beta}$ may be
determined, easily. Both of the latter generalizations introduced
by a deviation from the exponential function appears in the
weight function (and normalization) of canonical CSs. This idea
(which followed in PS CSs will be introduced later) is important,
due to the fact that Mandel parameter \cite{Mandel1,
Mandel2}(which demonstrates the statistics of any CSs) depends on
the normalization function of any CSs and its derivatives (see for
example \cite{kps}). So, deviation from exponential function,
implies the deviation from Poissonian statistics and forced the
CSs to possess the sub(or super) Poissonian statistics.
\subsection{Penson-Solomon (PS) generalized CSs:}
   The generalized
   CSs introduced by Penson and Solomon (PS) \cite{Penson} are as follows:
 \begin{equation}\label{ps1}
   |q, z\rangle_{\rm {PS}} = \N (q, |z|^2)^{-1/2} \sum_{n=0}^{\infty}
   \frac{q^{n(n-1)/2}}{\sqrt{n!}}z^n |n\rangle,
 \end{equation}
   where $\N (q, |z|^2)$ is a normalization function, $ 0\leq q \leq 1$.
   This definition is based on an entirely analytical prescription, in which
   the authors proposed the generalized exponential function obtained from the following differential equation:
 \begin{equation}\label{difeq}
  \frac {d \varepsilon(q, z)}{dz}=\varepsilon(q, qz) \Rightarrow
  \varepsilon(q, z)=\sum_{n=0}^\infty \frac {q^{n(n-1)/2}}{n!}z^n,
 \end{equation}
  where $\varepsilon(1, z)=\exp(z)$.
  This case is also based on the generalization of the exponential
  function (such as TC CSs).
   \subsection{ Barut-Girardello (BG) and Gilmore-Perelomov (GP) CSs for $su(1, 1)$ Lie algebra:}
    As an important generalized CSs one may refer to the
   BG CSs, defined for the discrete series
   representations of the group $SU(1, 1)$ \cite{bar-gir71}.
   These states can be realized in some physical systems such as
   the P\"oschl-Teller and infinite square well potentials.
   The BG CSs decomposed over the number-state bases as:
  \begin{equation}\label{BG1}
   |z, \kappa \rangle_{\rm {BG}} =\N_{\rm {BG}}(|z|^2)
  ^{-1/2}\sum_{n=0}^{\infty}\frac{z^n}{\sqrt{{n! \Gamma
   (n+2\kappa)}}}|n\rangle, \quad z \in \C \;,
 \end{equation}
   where $\N_{\rm{BG}}$ is a normalization constant
   and the label $\kappa$ takes the values $1, 3/2, 2, 5/2, ...$
   labels the $SU(1,1)$ representation being
   used.

 Similarly the GP CSs  defined as:
\begin{equation}\label{gilperCS}
  |z, \kappa \rangle_{\rm {GP}}=\N_{\rm{GP}}(|z|^2)^{-1/2}\sum_{n=0}^\infty\sqrt{\frac{\Gamma( n + 2\kappa
  )}{n!}}z^n|n\rangle, \qquad |z|<1,
\end{equation}
where again $\N_{\rm{GP}}$ is a normalization factor and $\kappa$
takes the same values, as in BG CSs.
\subsection{  $su(1, 1)$-Barut-Girardello (BG) CSs for Landau levels:}
   As an example more close to physics, it is well-known that
   the Landau levels  is directly related to quantum
   mechanical study of the motion of a charged and spinless particle on a flat
   plane in a constant magnetic field \cite{Gazeau, Landau} . Recently it
   is realized two distinct symmetries corresponding to these
   states, namely $su(2)$ and $su(1,1)$ in \cite{Fakhri}. The author
   showed that the quantum states of the Landau problem corresponding to the
   motion of a spinless charged particle on a flat
   surface in a constant magnetic field $\beta/2$ along $z-$axis may
   be obtained as:
\begin{equation}\label{LL-Fock}
   |n, m\rangle=\frac {e^{i m \varphi}}{\sqrt{2 \pi}}
   (\frac{r}{2})^{\frac{2\alpha +1}{2}}
   e^{-\beta r^2/8}L_{n,m}^{(\alpha,
   \beta)}(\frac{r^2}{4}),
\end{equation}
   where $0\leq \varphi \leq 2\pi$,
   $\alpha > -1, n\geq 0, 0 \leq m \leq n$ and $L_{n,m}^{(\alpha,
   \beta)}$ are the associated Laguerre functions.
   Constructing the Hilbert space spanned by $\HH :=\{|n, m\rangle\}_{n\geq 0, 0\leq m \leq
   n}$, there it is shown that the BG CSs associated to this
   system can be obtained as the following combination of the orthonormal basis:
 \begin{equation}\label{fakhri}
  |z\rangle_m=\frac{|z|^{(\alpha+m)/2}}{\sqrt
  {I_{\alpha+m}(2|z|)}}\sum_{n=m}^\infty \frac{z^{n-m}}
  {\sqrt{\Gamma(n-m+1)\Gamma(\alpha+n+1)}}|n, m\rangle ,
 \end{equation}
   where $I_{\alpha+m}(2|z|)$ is the modified Bessel function of
   the first kind \cite{Watson}. The states in (\ref{fakhri}) derived with the help of
   the lowering generator of the $su(1, 1)$ Lie algebra, the action of which defined by the relation:
 \begin{equation}\label{lower}
   K_-|n, m\rangle = \sqrt {(n+\alpha)(n-m)} |n-1, m\rangle,
   \qquad K_-|m, m\rangle=0.
 \end{equation}

\subsection{ Gazeau-Klauder generalized CSs (GK-CSs)}\label{GK}
    Adopting certain physical criteria rather than imposing selected
    mathematical requirements, Klauder and Gazeau  by
    reparametrizing the generalized CSs $|z\rangle$  in terms of a two independent
    parameters  $J$ and $\gamma$,
    introduced the generalized CSs $|J, \gamma \rangle$, known
    ordinarily  as Gazeau-Klauder CSs (we denoted them by GK-CSs) in the physical
    literature \cite{gazklau,  Klauder2001}. These states are
    explicitly defined by the expansion
 \begin{equation}\label{kps-jgama}
   |J, \gamma \rangle=\N(J)^{-1/2}
   \sum_{n=0}^{\infty}\frac{J^{n/2}e^{-ie_n\gamma}}{\sqrt{
   \rho(n)}}|n\rangle,
   \end{equation}
   where the normalization constant is given by
 $  \N(J)=\sum_{n=0}^{\infty}\frac {J^n}{\rho(n)}$,
   and $\rho(n)$ is a positive weight factor with $\rho(0)\equiv
   1$ by convention and the domains of $J$ and $\gamma$ are such that $J\geq 0$ and
   $-\infty < \gamma < \infty$. These states required to satisfy the
   following properties: (i) {\it continuity of labeling}: if
   $(J, \gamma)\rightarrow (J', \gamma')$ then,
   $\||J,\gamma \rangle -|J', \gamma'\rangle\|\rightarrow 0$, (ii)
   {\it resolution of the identity}: $\hat{I} = \int |J, \gamma \rangle
   \langle J, \gamma |d \mu (J, \gamma)$ as usual and two extra
   properties: (iii) {\it temporal stability}: $ \exp (-i \hat{H} t) |J,
   \gamma \rangle = |J,\gamma + \omega t \rangle$ and (iv) {\it the action
   identity}: $ H=\langle J, \gamma |\hat{H}|J, \gamma \rangle=\omega
   J$, where $H$ and $\hat{H}$ are classical and quantum mechanical
   Hamiltonians of the system, respectively. It must be understood
   that the forth condition forced the generalized CSs to have the
   essential property: {\it "the most classical quantum states"},
   but now in the sense of {\it "energy"} of the dynamical system, in the
   same way that the canonical CSs is a quantum state which its
   position and momentum expectation values obey the classical
   orbits of harmonic oscillator in phase space.
   It must also be noted that for the second criteria,  Gazeau and Klauder  defined
 \begin{eqnarray}\label{Res-I2}
    \int |J, \gamma \rangle \langle J, \gamma| d\mu(J, \gamma)
   & = &
   \lim_{\Gamma \rightarrow \infty}
   \frac{1}{2\Gamma} \int _{-\Gamma}^{\Gamma} d\gamma
   \int_0^R dJ {\N}(J)\varrho(J) |J, \gamma \rangle \langle J,
   \gamma| \nonumber\\
   & = &
 \sum_{n=0}^\infty  |n \rangle \langle n| \int_0^R dJ \varrho(J) \frac{J^n}
{\rho(n)} = \sum_{n=0}^\infty  |n \rangle \langle n| = \hat{I} \;.
\end{eqnarray}
 This condition finally resulted in the moment problem as follows:
 \begin{equation}\label{GK-mom}
     \int_0^R dJ J^n \varrho(J) =\rho(n)\cdot
 \end{equation}
   In Eq. (\ref{kps-jgama}) the kets ${|n\rangle}$
   are the eigen-vectors of the Hamiltonian $\hat{H}$,  with the
   eigen-energies $E_n$,
 \begin{equation}\label{kps-14}
  \hat{H}|n\rangle=E_n|n \rangle \equiv \hbar \omega e_n |n
  \rangle \equiv e_n|n\rangle, \quad \hbar \equiv 1,
  \quad \omega\equiv 1, \quad  n=0,1, 2, ... \;,
 \end{equation}
   where the re-scaled spectrum $e_n$, satisfied the inequalities
   \begin{equation} \label{}
        0=e_0 < e_1 < e_2 < \cdots < e_n < e_{n+1} < \cdots \;.
   \end{equation}
   The action identity {\it uniquely} specified $\rho(n)$ in terms of
   the eigen-values of the Hamiltonian $\hat{H}$
 \begin{equation}\label{kps-en}
   \rho(n)=\Pi_{k=1}^n e_k\equiv[e_n]! \; .
 \end{equation}
   As an example, for the shifted Hamiltonian of harmonic oscillator
   we have the canonical CSs in the language of Gazeau-Klauder, denoted by $|J, \gamma \rangle_{CCS}$:
 \begin{equation}\label{kpsCC-16}
   |J, \gamma \rangle_{_{CCS}} =
   e^{-{J/2}}\sum_{n=0}^{\infty}\frac{J^{n/2}e^{-in\gamma}}{\sqrt{n!}}
   |n\rangle.
 \end{equation}
   Eq. (\ref {kps-en}) obviously states that the functions $\rho(n)$ is directly
   related to the spectrum of the dynamical system. So every
   Hamiltonian {\it uniquely} determined the associated CS, although the
   inverse is not true \cite{Fern'andez, rokntvs}.
  %
\subsubsection{ Analytical representations of Gazeau-Klauder generalized CSs (GKCSs)}\label{GKCS}
    El Kinani and Daoud, in a series of papers
       \cite{ElKinani2001b, ElKinani2002, ElKinani20022, ElKinaniand_Daoud, Elkinani2003}
    imposed a minor modification on the GK-CSs  in (\ref{kps-jgama})
    via generalizing the Bargman representation for the standard
    harmonic oscillator \cite{Bargman1961}.
    Following the path of Gazeau-Klauder, the authors
    introduced the {\it analytical representations of GK-CSs},
    denoted by us as GKCSs:
\begin{equation}\label{GKED}
    |z, \alpha \rangle \doteq \N(|z|^2)^{-1/2}
    \sum_{n=0}^{\infty}\frac{z^{n}e^{-i \alpha e_n}}{\sqrt{
    \rho(n)}}|n\rangle,  \qquad  z \in \C, \qquad \alpha \in \R,
    \end{equation}
  where the normalization constant is given by
    $\N(|z|^2)=\sum_{n=0}^{\infty}\frac {|z|^{2n}}{\rho(n)}$ and the function $\rho(n)$ can be expressed in terms of
     the eigen-values as $\rho(n)=[e_n]!$.
   Briefly speaking,
  they replaced $-\infty < \gamma < \infty$ and $J>0$ in (\ref{kps-jgama})
   by $\alpha \in \R$ and $z \in \C$, respectively.
   We must emphasize the main difference between the
   GKCSs presented in (\ref {GKED}) and GK-CSs in (\ref
   {kps-jgama}) in view of the significance and the role of $\gamma$ and
   $\alpha$, particularly in the integration procedure, in order to
   establish the resolution of the unity.
   According to their proposal for this purpose it is required to find an appropriate
   positive measure $d \lambda(z)$ such that the following integral satisfied:
 \begin{equation}\label{RI-GKED}
     \int_0^{R}|z, \alpha\rangle \langle z, \alpha|d \lambda(z)=
     \sum_{n=0}^\infty|n\rangle\langle n|= \hat{I},
     \qquad 0 < R \leq \infty.
 \end{equation}
   Inserting (\ref {GKED}) in (\ref{RI-GKED}), writing $z= x e^{i\theta}$ and
   then expressing the measure as:
 \begin{equation}\label{measure}
   d \lambda(z)=d \lambda(|z|^2)=\pi \N(x^2) \sigma(x^2) x dx d \theta,
 \end{equation}
  performing the integration over $\theta \in [0, 2\pi]$, the over-completeness
  relation (\ref {RI-GKED}) finally
  boils down to the moment problem (see \cite{kps} and Refs. therein, especially \cite{Akhiezer}):
 \begin{equation}\label{mom}
   \int_0 ^R  x^ n  \sigma(x) dx = \rho (n).
 \end{equation}
  Note that while we have integrated on $\gamma$ in (\ref{Res-I2}), this is not hold for $\alpha$ parameter
  in the procedure led to (\ref{mom}).
\section{The unification methods}
  In what follows two approaches will be presented which by them I
  try to come back to the one (or more) of the three customary
  generalization methods of constructing the MP CSs we revisited  in section 2, as well as some others.
  As  mentioned
  in the introduction, while  these {\it "unification methods"} have been established,  {\it "some
  new classes of generalized CSs"} will be obtained.
  Indeed, we duplicate the number of them through introducing the dual family
  associated with each classes of them.

\subsection{The first approach: the nonlinear CSs method}\label{sec-relat}
   This approach is basically based on the
   conjecture that all classes of the generalized CSs outlined in section 2 can be interpreted as nonlinear CSs.
 \vspace{4 mm}\\
   {\it Example $1\;$ KPS and ML  CSs as nonlinear CSs}
   \vspace{2 mm}\\
   To start with, we demonstrate the relation between the KPS (and therefore ML CSs) and
   nonlinear  CSs.  Looking at the KPS CSs in (\ref{kps-cs}) and using   (\ref{findf}) yields
   the nonlinearity function as:
 \begin{equation}\label{kps-20}
   f_{_{\rm KPS}}(\hat{n})=\sqrt{\frac{\rho(\hat{n})}{\hat{n}\rho(\hat{n}-1)}},
 \end{equation}
   which provides simply a bridge
   between KPS and  nonlinear CSs. As a special case when
   $\rho(n)=n!$, i.e. the canonical CSs, we obtain $f_{\rm KPS}(n)=1$.

   From Eq. (\ref {kps-en}) $e_n$
   can easily be found in terms of $\rho(n)$:
 \begin{equation}\label{kps-21}
   e_n=\frac{\rho(n)}{\rho(n-1)}.
 \end{equation}
   Remembering that $e_n$s are the eigen-values of the Hamiltonian, it will be obvious
   that neither every $\rho(n)$ of KPS CSs nor every
   $f_{\rm KPS}(n)$ of nonlinear CSs are physically acceptable, when the dynamics
   of the system (Hamitonian) is specified. Using Eqs. (\ref
   {kps-20}) and (\ref {kps-21}) we get
 \begin{equation}\label{fn-en}
   f_{_{\rm KPS}}(n)=\sqrt{\frac{e_n}{n}}   \Leftrightarrow  e_n=n(f_{_{\rm KPS}}(n))^2.
 \end{equation}
   Now by using (\ref{kps-20}) we are able to find $f(\hat{n})$ for
   all sets of KPS CSs (on the whole plane and on a unit disk) and then  the deformed
   annihilation and creation operators $A=af_{KPS}(\hat{n})$ and
   $A^\dag=f_{KPS}^\dag(\hat{n})a^\dag$ may easily be obtained. Returning to
   the above descriptions, the operators $A$ and $A^\dag$ satisfy the
   following relations :
 \begin{equation}\label{Adag1}
   A|n\rangle =\sqrt{e_n} |n-1\rangle,
 \end{equation}
 \begin{equation}\label{Adag2}
   A^\dag|n\rangle =\sqrt{e_{n+1}} |n+1\rangle,
 \end{equation}
 \begin{equation}\label{AAdag3}
   [A, A^\dag]|n \rangle =( e_{n+1}- e_n) |n\rangle, \quad [A,
   \hat{n}]=A, \quad [A^\dag, \hat{n}] = -A^{\dag}.
 \end{equation}
   Also note that $A^{\dag}A |n\rangle=e_n|n\rangle$, not equal to $n|n\rangle$ in general.
   With these results in mind, it would be obvious that Eq. (\ref{fn-en})
   is not consistent with the relations (\ref{hamilt})
   and (\ref{hamilt-f}) for the Hamiltonian. A closer look at Eq. (\ref{fn-en})
   which is a consequence of imposing the action
   identity on the Hamiltonian  of the system leads us to obtain
   a new form of the Hamiltonian for the nonlinear CSs as
 \begin{equation}\label{kps-23}
   \hat{H} |n\rangle = n f^2(n)|n\rangle    \Leftrightarrow      \hat{H}=\hat{n}f^2(\hat{n})=A^\dag A.
 \end{equation}
   Consequently, this form of the Hamiltonian  may be considered as
   {\it ''normal-ordered''} of the Man'ko {\etal} Hamiltonian $H_{\rm M}$, introduced
   in Eq. (\ref{hamilt}),
    $  \hat{H}=\;\;:\hat{H}_M:.$
   Therefore the associated Hamiltonian for the KPS CSs
   can be written as
 \begin{equation}\label{kps-25}
   \hat{H}_{_{\rm KPS}}=\hat{n}(f_{_{\rm KPS}}(\hat{n}))^2=\frac{\rho(\hat{n})}{\rho(\hat{n}-1)}.
 \end{equation}
   Comparing Eqs. (\ref {hamilt}) and (\ref {hamilt-f}) with the
   Eqs. (\ref {fn-en}), (\ref {kps-23}),
   implies that if we require that the KPS and nonlinear CSs
   possess the action identity property, the associated Hamiltonian when
   expressed in terms of ladder operators must be reformed in
   normal-ordered form.

   In summary our considerations enable one to obtain $f$-deformed
   annihilation and creation operators (and also as we will see in later sections,
   the generalized displacement operators) as well as the Hamiltonian for all
   sets of ${|z\rangle_{_{\rm KPS}}}$ discussed in Ref. \cite{kps} (on the whole plane and on a unit disk), after
   demonstrating that for each of them there exist a special nonlinearity
   function $f(\hat{n})$ (for the detail of results see  \cite{Roknizadeh2004}).
\vspace{4 mm}\\
   {\it Example $2\;$ HG  CSs as nonlinear CSs}
   \vspace{2 mm}\\
    Applying the presented formalism in this section  on the HG  CSs in (\ref{GHCS}) yields  the
    following nonlinearity function
\begin{equation}\label{nl-hgcs}
  \fl {}_p f _q(\hat n) ={} _p f _q(\alpha_1, \cdots \alpha_p; \beta_1,
  \cdots \beta_q; \hat n) = \sqrt{ (\hat n-1) \frac{(\beta_1+\hat n-1) \cdots
 (\beta_q+\hat n-1)}{(\alpha_1+\hat n-1) \cdots (\alpha_p+\hat
  n-1)}} \; ,
\end{equation}
    and the dynamical  Hamiltonian corresponding to such systems as
\begin{equation}\label{hami-hgcs}
  \hat H (\hat{n})=\hat n _p f^2 _q(\hat n) = \hat n (\hat n-1) \frac{(\beta_1+\hat n-1) \cdots
 (\beta_q+\hat n-1)}{(\alpha_1+\hat n-1) \cdots (\alpha_p+\hat
  n-1)}\; ,
\end{equation}
which is clearly $\hat n$-dependent.
\vspace{4 mm}\\
  {\it Example $3\;$   TC CSs as nonlinear CSs}
  \vspace{2 mm}\\
The Tricomi (TC) CSs of the first kind introduced in
(\ref{TC-1}). For the nonlinearity functions responsible to these
states, as well as the dynamical Hamiltonians of such system we
get
\begin{equation}\label{TC-1-nl}
\fl f_{\rm TC}^{(1)} (\hat n)= \sqrt{ \frac {2}{\sqrt p}\; \frac
{\psi(\frac{\hat n+1}{2};\frac 1 2;\frac{1}{4p})
}{\psi(\frac{\hat n}{2};\frac 1 2;\frac{1}{4p} )}}, \qquad \hat
H^{(1)} (\hat n) = \hat n
 \frac {2}{\sqrt p}\; \frac {\psi(\frac{\hat n+1}{2};\frac 1
2;\frac{1}{4p}) }{\psi(\frac{\hat n}{2};\frac 1 2;\frac{1}{4p}
)}\; .
\end{equation}
Similarely for the second kind of Tricomi's CSs defined in
(\ref{TC-2}) we find
\begin{equation}\label{TC-2-nl}
 \fl f_{\rm TC}^{(2)} (\hat n) = \sqrt{ \beta \;  \frac {\psi(\hat n+1;
 \hat n+2-\lambda; \beta) }{\psi(\hat n; \hat n+1-\lambda;
 \beta)}}, \qquad \hat H^{(2)}  (\hat n) = \hat n \beta \; \frac
 {\psi(\hat n+1; \hat n+2-\lambda; \beta) }{\psi(\hat n; \hat
 n+1-\lambda; \beta)}\; .
\end{equation}
as the nonlinearity function and the Hamiltonian which are clearly
$\hat n$-dependent. Therefore, again setting  $f_{\rm TC}^{(1)}$
and $f_{\rm TC}^{(2)}$ in (\ref{nonl-annih}) one can easily gets
the ladder operators responsible to such systems of CSs and both
of them can be obtained by annihilation operator definition.
   \vspace{4 mm}\\
   {\it Example $4\;$ PS generalized CSs as nonlinear CSs}
   \vspace{2 mm}\\
   As another  example we
   can simply deduce the nonlinearity function for the generalized
   CSs introduced by Penson and Solomon (PS) \cite{Penson}.
   As it may be observed,  in the construction of PS CSs (see Eq. (\ref{ps1})),
    the authors have not used the ladder(or displacement) operator
   definition for their states.
   But the present  formalism helps one to find
   the nonlinearity function (and so the  annihilation
   and creation operators evolve in these states) as
 \begin{equation}\label{ps2}
   f_{_{\rm PS}}(\hat{n})= q ^ {1-\hat{n}}.
 \end{equation}
   Therefore the method enables one to reproduce them
   through solving  the eigen-value equation
   $A|z, q\rangle_{_{PS}}=a q^{1-\hat{n}}|z, q\rangle_{_{PS}}= z |z, q\rangle_{_{PS}}$,
   in addition to introducing a $\hat{n}$-dependent Hamiltonian describing the dynamics of the
   system:
 \begin{equation}\label{psham}
   \hat{H}_{\rm {PS}}(\hat{n})=\hat{n} q ^{2(1-\hat{n})}.
 \end{equation}

   \vspace{4 mm}
   {\it Example $5\;$ BG CSs for $su(1, 1)$ Lie algebra as nonlinear CSs}
   \vspace{2 mm}\\
   Using (\ref{findf}) for BG CSs  in (\ref {BG1}), one may find
   \begin{equation}\label{BG11}
   f_{\rm {BG}}(\hat{n})=\sqrt {\hat{n}+2\kappa -1},
   \quad n=0,1,2,... \;.
 \end{equation}
 So, an $\hat{n}$-dependent Hamiltonian describing the dynamics of the
   system immediately obtained  as:
 \begin{equation}\label{BG2}
  \hat{H}_{\rm {BG}}(\hat{n}) =\hat{n}(\hat{n}+2\kappa-1),
 \end{equation}
   with    $\hat{H}_{\rm {BG}}|n\rangle = e_n |n\rangle$; When $\kappa=3/2$ and
   $\kappa=\lambda+\eta$ ($\lambda$ and $\eta$ are two parameter
   characterize the P\"{o}sch-Teller potential with $\kappa=[(\lambda + \eta+1)/2]>3/2$)
   we obtain the infinite square well and P\"{o}schl-Teller potentials, respectively
   \cite{Antoine2001}.
   Therefore  the dynamical group
   associated with these two potentials is the $SU(1, 1)$ group have been established.
   Also, if one takes into account the action of
   $A=af_{\rm {BG}}(\hat{n}), A^{\dag}=f_{\rm {BG}}(\hat{n})a^{\dag}$ and $[A,
   A^{\dag}]$ on the states $|\kappa, n\rangle$ we obtain
 \begin{equation}\label{BG3}
    A|\kappa, n\rangle=\sqrt{n(n+2\kappa-1)}|\kappa, n-1\rangle,
 \end{equation}
 \begin{equation}\label{BG4}
    A^{\dag}|\kappa, n\rangle= \sqrt{(n+2\kappa)(n+1)}|\kappa,
    n+1\rangle,
 \end{equation}
 \begin{equation}\label{BG5}
    [A, A^{\dag}]|\kappa, n\rangle=(n+\kappa)|\kappa, n\rangle.
 \end{equation}
   We may conclude that the generators of $su(1, 1)$ algebra $\hat L_-,\hat L_+,\hat L_{12}$
   can be expressed in terms of the
   deformed annihilation and creation operators including their commutators
   such that
 \begin{eqnarray}\label{BG6}
  & & \hat L_- \equiv \frac{1}{\sqrt 2}A=\frac{1}{\sqrt 2}a
    f_{\rm {BG}}(\hat{n}), \nonumber\\
  & &  \hat L_+ \equiv \frac{1}{\sqrt 2}A^{\dag}=\frac{1}{\sqrt 2}
    f_{\rm {BG}}(\hat{n})a^{\dag}, \nonumber\\
   & & \hat L_{12}\equiv \frac{1}{2}[A, A^{\dag}].
 \end{eqnarray}

   Note that the presented  approach not only recovers the results of Ref.
   \cite{Antoine2001} in a simpler and clearer manner, but also it gives the
   explicit form of the operators ${\hat L_-,\hat L_+,\hat L_{12}}$ as some
   {\it "intensity dependent"} operators in consistence with the
   Holstein-Primakoff single mode realization of the $su(1, 1)$
   Lie algebra \cite{Gerry}.
   Obviously, similar discussion can be followed for the GP representations of $SU(1,1)$ group.
  \vspace{4 mm}\\
  {\it Example $6\;$  $su(1, 1)$-BG CSs for Landau levels as nonlinear CSs}
  \vspace{2 mm}\\
    BG CSs associated with  algebra $su(1, 1)$ for Landau
    levels have been derived in Ref. (\ref{fakhri}).
    A deeply inspection to these states,
    in comparison to the states in (\ref {BG1}) shows a little
    difference, in view of the lower limit of the summation sign in the
    former Eq. (\ref{fakhri}). But this situation is similar to
    the states known as photon-added CSs \cite{Agarval-tara} in the
    sense that both of them are combinations of Fock space,
    with a cut-off in the summation from
    below. This common feature leads one to go on with the same
    procedure that has been done already to yield the
    nonlinearity function of the photon-added CSs in \cite{Siv1999}.
    Let define the deformed annihilation operator and the
    nonlinearity function similar to photon-added CSs as
 \begin{equation}\label{defAf}
   A=f(\hat{n})a, \qquad f(n)=\frac{C_n}{\sqrt{n+1}C_{n+1}},
 \end{equation}
   where  the states (\ref{nonl-cs}) have been used (replacing $|n\rangle$ by $|n, m\rangle$)
   as the eigen-states of the new annihilation operator defined in Eq. (\ref{defAf}).
   Upon these considerations one can calculate the nonlinearity
   function for the $su(1, 1)$-BG CSs related to Landau level as follows:
 \begin{equation}\label{}
   f_{_{\rm LL}}(\hat{n})=\frac{(\hat{n}-m+1)(\hat{n}+\alpha+1)}{\sqrt{\hat{n}+1}}.
 \end{equation}
   Hence again the ladder operators corresponding to this system may
   be  easily obtained with the usage of (\ref{nonl-annih}).
   Similar to previous cases, the $\hat{n}$-dependent Hamiltonian describing the dynamics of the
   system obtained  as:
\begin{equation}\label{}
   \hat H(\hat{n}) = \hat{n} \; \frac{(\hat{n}-m+1)^2(\hat{n}+\alpha+1)^2}{{\hat{n}+1}}.
 \end{equation}
Two points must be emphasized here, at the end of this subsection.
The first is that this first approach (as well as the second one
will be explained later) does not works well for GK-CSs and GKCSs,
so   attention will be paied to these cases in  future sections 5
and 6. The second is that, although  the deformed annihilation
and creation operators have not been introduced explicitly in
neither of the foregoing examples, finding these operators when
the nonlinearity functions have been introduced is very clear
using Eq. (\ref{nonl-annih}).

Therefore, to this end all of the discussed mathematical physics
CSs may be reconstructed by {\it "annihilation operator
definition"}, i.e. $A_f |z,f\rangle_{\rm MP} = z |z,f\rangle
_{\rm MP}$, where $A_f$ and $|z,f\rangle _{\rm MP}$ are the
 $f-$deformed annihilation operators associated with each class  of MP CSs, and the MP CSs
introduced in this subsection, respectively. So, the algebraic
definition of MP CSs has been established, clearly. Besides, we
arrive at an $n-$dependent Hamiltonian related to each set of MP
CSs, which describes the dynamics of the corresponding system.
The third definition, the group theoretical definition and the
generalized displacement operator, will be discussed in section 4.
Altogether in this manner {\it we obtain the lost ring between the
MP CSs and the three generalization methods, successfully}.
\subsection{The second approach: the general setting of mathematical physics method}\label{sec-genset}

  The primary object for this discussion will be an abstract Hilbert space $\HH$.
  Let $T$ be an operator on this space with the properties
  \cite{AliRokTav}
\begin{enumerate}

\item $\hat T$ is densely defined and closed; we denote its domain by $\D(T)$.

\item $\hat T^{-1}$ exists and is densely defined, with domain $\D(T^{-1})$.

\item The vectors $\phi_n\in\D(\hat  T)\cap\D(\hat  T^{-1})$ for all $n$ and there exist non-empty
      open sets $\D_{\hat T}$ and $\D_{\hat  T^{-1}}$ in $\mathbb C$ such that
      $\eta_z \in \D(\hat  T), \forall z \in \D_{\hat T}$ and $\eta_z \in \D(\hat  T^{-1}), \forall z \in \D_{\hat T^{-1}}$.
\end{enumerate}
Note that condition (1) implies that the operator $\hat T^\ast
\hat T=\hat F$ is self adjoint.

Let
\begin{equation}\label{dual-transf}
\phi_n^F    := \hat   T^{-1}\phi_n\; ,  \qquad \phi_n^{F^{-1}} :=
\hat  T\phi_n\; , \qquad n =0, 1, 2, \ldots , \infty\; ;
\end{equation}
we define the two new Hilbert spaces:
\begin{enumerate}

  \item $\HH_F$, which is the completion of the set $\D( \hat T)$ in the scalar
product
\begin{equation}\label{F-space}
  \langle f|g\rangle_F  =  \langle f|\hat T^\ast \hat T g\rangle_\HH
     = \langle f|\hat  F g\rangle_\HH.
\end{equation}
The set $\{\phi_n^{F}\}$ is orthonormal in $\HH_F$ and the map
$\phi \longmapsto \hat T^{-1}\phi ,\; \phi \in \D(\hat  T^{-1})$
extends to a {\em unitary} map between $\HH$ and $\HH_{F}$. If
both $\hat T$ and $\hat T^{-1}$ are bounded, $\HH_F$ coincides
with $\HH$ as a set. If $\hat T^{-1}$ is bounded, but $\hat T$ is
unbounded, so that the spectrum of $\hat F$ is bounded away from
zero, then $\HH_F$ coincides with $\D(\hat T)$ as a set.

\medskip

  \item $\HH_{F^{-1}}$, which is the completion of $\D(\hat T^{\ast -1})$ in the
scalar product
\begin{equation}
  \langle f|g\rangle_{F^{-1}}  =   \langle f|\hat T^{-1}\hat T^{\ast-1}
  g\rangle_\HH
  = \langle f|\hat F^{-1}g\rangle_\HH.
\label{F-inv-space}
\end{equation}
  The set $\{\phi_n^{F^{-1}}\}$ is orthonormal in $\HH_{F^{-1}}$
and the map $\phi\longmapsto \hat T\phi ,\; \phi \in \D(\hat T)$
extends to a {\em unitary} map between $\HH$ and $\HH_{F^{-1}}$.
If the spectrum of $\hat F$ is bounded away from zero, then $\hat
F^{-1}$ is bounded and one has the inclusions
\begin{equation}\label{gelf-triple}
  \HH_F\subset\HH\subset\HH_{F^{-1}}\; .
\end{equation}
\end{enumerate}

The spaces $\HH_{F}$ and $\HH_{F^{-1}}$ will refer to as a {\em
dual pair} and when (\ref{gelf-triple}) is satisfied, the three
spaces $\HH_{F}$, $\HH$ and $\HH_{F^{-1}}$ will be called a {\em
Gelfand triple\/} \cite{Gelfand}. (Actually, this is a rather
simple example of a Gelfand triple, consisting only of a triplet
of Hilbert spaces \cite{antintr}).

On $\HH$ we take the operators $a, a^\dag, \hat n = a^\dag a$:
\begin{equation}\label{osc-alg1}
a\phi_n=\sqrt{n}\phi_{n-1},\quad
a^\dag\phi_n=\sqrt{n+1}\phi_{n+1},\quad \hat n\phi_n=n\phi_n \; .
\end{equation}
These operators satisfy:
\begin{equation}\label{osc-alg2}
  [a,a^\dag]=\hat I,\quad [a,\hat n]=a,\quad [a^\dag, \hat n]=-a^\dag\; .
\end{equation}
On $\HH_F$ we have the transformed operators:
\begin{equation}\label{osc-alg3}
  \af=\hat T^{-1}a \hat T,\quad a^\dag_F= \hat T^{-1}a^\dag \hat T,\quad \hat n_F=T^{-1}\hat n \hat T\; .
\end{equation}
These operators satisfy the same
 commutation relations as $a, a^\dag $ and  $\hat n$ :
\begin{equation}\label{osc-alg4}
  [\af,a^\dag_F]=\hat I,\quad [\af,\hat n_F]=\af,\quad [a^\dag_F, \hat n_F]=-a^\dag_F\; .
\end{equation}
Also on $\HH_F$
\begin{equation}\label{osc-alg5}
\af\phi_n^F=\sqrt{n}\phi_{n-1},\quad
a^\dag_F\phi_n^F=\sqrt{n+1}\phi_{n+1}^F,\quad \hat n
_F\phi_n^F=n\phi_n^F\; .
\end{equation}
 Clearly, considered as operators on
$\HH_F$, $\af$ and $a^\dag_F$ are adjoints of each other and
indeed they are just the unitary transforms on $\HH_F$ of the
operators $a$ and $a^\dag$ on $\HH$. On the other hand, if we
take the operator $\af$, let it act on $\HH$ and look for its
adjoint on $\HH$ under this action, we obtain by (\ref{osc-alg3})
the operator $a^\sharp=\hat T^\ast a^\dag \hat T^{\ast-1}$ which,
in general, is different from $a^\dag_F$ and also
$[\af,a^\sharp]\neq \hat I$, in general. In an analogous manner,
we shall define the corresponding operators $\afy ,  \afdy, \hat
n_F$, on $\HH_{F^{-1}}$ with their related actions on
$\{\phi_n^{F^{-1}}\}_{n=0}^\infty$.

   Therefore, three unitarily equivalent sets of operators have been obtained: $a, a^\dag ,  \hat n$, defined on $\HH$,
$\af, \afd ,  \hat n_F$, defined on $\HH_F$ and $\afy, \afdy ,
\hat n_{F^{-1}}$ defined on $\HH_{F^{-1}}$. On their respective
Hilbert spaces, they define under commutation the standard
oscillator Lie algebra. On the other hand, if they are all
considered as operators on $\HH$, the algebra generated by them
and their adjoints on $\HH$ (under commutation) is, in general,
very different from the oscillator algebra and could even be an
infinite dimensional Lie algebra.

Writing $A=\af$, $A^\dag=a^\sharp$, both considered as operators
on $\HH$, if they satisfy the relation
\begin{equation}\label{def-osc-alg}
  AA^\dag-\lambda A^\dag A=C(\hat n),
\end{equation}
where $\lambda\in\R_\ast^+$ is a constant and $C(\hat n)$ is a
function of the operator $\hat n$, then the three operators $A$,
$A^\dag$, $\hat H=(1/2)(AA^\dag+A^\dag A)$ or according to
(\ref{kps-23}) its normal ordered form, are said to generate a
{\it "generalized oscillator algebra"} or {\it "deformed
oscillator algebra"} \cite{borzov95}. Note that on $\HH$, $A$ and
$A^\dag$ are adjoints of each other.

\subsection{Construction of generalized CSs}\label{sec-CS-cons}
Considering   a  somewhat more general notation and writing the
canonical CSs as,
\begin{equation}\label{CCS-gen}
  \vert z \rangle = \eta_z = \N(|z|^2)^{-1/2}\sum_{n=0}^\infty\frac{z^n}{\sqrt{n!}}\;\fn,\quad
 \forall z\in\C\; ,
\end{equation}
 defined as vectors in an abstract (complex, separable) Hilbert
 space $\HH$, for which the vectors $\fn$ form an orthonormal basis
  $\langle\phi_n|\phi_m\rangle_{\HH}=\delta_{nm},\;\; n,m=0,1,2,\cdots,\infty\;$.
 Then consider the vectors
\begin{equation}\label{F-CS}
  \eta_z^F= \hat T^{-1}\eta_z=\N(|z|^2)^{-1/2}\sum_{n=0}^\infty\frac{z^n}{\sqrt{n!}}\;\phi_n^F
\end{equation}
on $\HH_F$. These are the images of the $\eta_z$ in $\HH_F$ and
are the normalized canonical CSs on this Hilbert space (recall
that the vectors $\phi_n^F$ are orthonormal in $\HH_F$).
Similarly, define the vectors
\begin{equation}\label{F-inv-CS}
  \eta_z^{F^{-1}}= \hat T \eta_z=\N(|z|^2)^{-1/2}\sum_{n=0}^\infty \frac
    {z^n}{\sqrt{n!}}\;\phi_n^{F^{-1}}\; ,
\end{equation}
as the canonical CSs $\eta_z$ unitarily transported from $\HH$ to
$\HH_{F^{-1}}$.

  Let me now consider the  $\eta_z^F$ as being vectors
in $\HH$ and similarly the vectors $\eta_z^{F^{-1}}$ also as
vectors in $\HH$. To what extent can we then call them
(generalized) CSs? Specifically, we would like to find an
orthonormal basis $\{\psi_n\}_{n=0}^\infty$ in $\HH$ and a
transformation $w=f(z)$ of the complex plane to itself such that:
\begin{itemize}
  \item[$(a)$] one could write,
\begin{equation}\label{transf-nlCS}
  \eta_z^F=\zeta_w=\N^\prime (|w|^2)^{-1/2}\; \Omega (w)\sum_{n=0}^\infty\frac{w^n}{\sqrt{[x_n!]}}\;\psi_n\; ,
\end{equation}
where $\N^\prime$ is a new normalization constant, $\Omega (w)$
is a phase factor and $\{x_n\}_{n=1}^\infty$ is a sequence of
non-zero positive numbers, to be determined;

\medskip

\item[$(b)$] there should exist a measure $d\lambda(\rho)$ on $\R^+$, such
that with respect to the measure
$d\mu(w,\overline{w})=d\lambda(\rho)\; d\vartheta$ (where $w=\rho
e^{i\vartheta}$) the resolution of the identity,
\begin{equation}\label{resolid}
  \int_\D
  |\zeta_w\rangle\langle\zeta_w|\N^\prime(|w|^2)d\mu(w,\overline{w})= \hat I\; ,
\end{equation}
would hold on $\HH$ (as is the case with the canonical coherent
states). Here again, $\D$ is the domain of the complex plane,
$\D=\{w\in \C \;\vert\; \vert w\vert < L\}$, where $L^2
=\lim_{n\to\infty}x_n$.
\end{itemize}
A general answer to the above question may be hard to find. But
several classes of examples will be presented below, all
physically motivated, for which the above construction can be
carried out. These include in particular all the so called
nonlinear as we will observe in this tutorial and the deformed
and squeezed CSs (see for detail our work \cite{AliRokTav}),
which appear so abundantly in the quantum optical and physical
literature \cite{Manko1997, Odzijewicz, Simon}.

 Whenever the two sets of vectors $\{\eta_z^F\}$ and $\{\eta_z^{F^{-1}}\}$
 form CSs families in the above sense, we shall call
 them a {\it dual pair}.

 \subsection{Examples of the general construction of CSs}\label{sec-ex}

 \subsubsection{ Photon-added and binomial states as bases}

 Let $\hat T$ be an operator such that $\hat T^{-1}$ has the form
\begin{equation}\label{ex1T}
\hat  T^{-1}=e^{\lambda a^\dag}G(a),
\end{equation}
where $\lambda\in\R$ and $G(a)$ is a function of the operator $a$
such that $\hat T$ and $\hat T^{-1}$ satisfy the postulated
conditions (1)-(3) of Section \ref{sec-genset}. We can compute
the two transformed operators $\af$ and $a^\dag_F$ on $\HH_F$
($\hat F=\hat T^\ast \hat T=e^{-\la
a}G(a^\dag)^{-1}G(a)^{-1}e^{-\la a^\dag}$) to be:
\begin{equation}\label{ex1-adj3}
\af  = \hat  T^{-1}a \hat T=a-\la \hat I\; , \qquad \afd = \hat
T^{-1}a^\dag \hat T=G(a-\la \hat  I)a^\dag G(a-\la \hat I)^{-1}\;
.
\end{equation}
Thus, since $a$ commutes with $G(a-\la \hat  I)$, we obtain
  $[a^{ }_F,a^\dag_F] = \hat I$, as expected. The two operators $A=\af$ and $A^\dag=\hat T^\ast a^\dag
 \hat T^{\ast^{-1}}$, defined on $\HH$, are
\begin{equation}\label{ex1-adj4}
  A=a-\la \hat  I,\quad A^\dag=a^\dag-\la \hat  I,
\end{equation}
which of course are adjoints of each other. Moreover, in this case
  $[A,A^\dag]= \hat I\; ,$
so that the oscillator algebra remains unchanged.

  By some similar procedures,  we get the corresponding operators,
  \begin{equation}\label{ex1-adj7}
  \afy = \hat Ta \hat T^{-1}=a+\la \hat  I\; , \qquad
  \afdy  = \hat  Ta^\dag \hat T^{-1}=G(a)^{-1}a^\dag G(a)\; ,
  \end{equation}
on $\HH_{F^{-1}}$. Once again we obtain $[\afy,\afdy]= \hat I$ and
similarly for the operator $A'=\afy=a+\la \hat I$ and its adjoint
$A'^\dag=a^\dag+\la \hat I$ on $\HH$.

Let now define the vectors
\begin{equation}\label{phot-add-bin}
  \phi^F_n=\hat T^{-1}\phi_n=e^{\la a^\dag}G(a)\phi_n\; ,
\end{equation}
which form an orthonormal set in $\HH_F$, and build the
corresponding canonical CSs
\begin{equation}\label{phot-add-CS}
  \eta^F_z = =e^{\la
  a^\dag}G(a)\eta_z = \N(|z|^2)^{-1/2}\sum_{n=0}^\infty\frac{z^n}{\sqrt{n!}}\phi_n^F\; ,
\end{equation}
on $\HH_F$. Considering these as vectors in $\HH$, one can see
that
\begin{equation}\label{phot-add-bin2}
a \eta^{F}_{z}= (z+\la)\eta^{F}_{z}.
\end{equation}
Thus, up to a constant factor, $\eta^F_z$ is just the canonical
CSs on $\HH$ corresponding to the point $(z+ \lambda)\in \C$ such
that the following holds for $\eta_z^F$ in (\ref {phot-add-CS})
$$
 \eta_z^{F}= C(\lambda , z )\sum_{n=0}^\infty\frac{(z+\lambda)^n}{\sqrt{n!}}\;\fn\; ,
$$
where the constant $C(\lambda , z )$ can be computed by going
back to (\ref{phot-add-CS}). Thus, one gets $C(\lambda , z ) =
G(z)e^{- \frac {\vert z \vert^2}2}$ and finally
\begin{equation}\label{phot-add-bin3}
 \eta_z^F = G(z)e^{- \frac {\vert z \vert^2}2}\;\sum_{n=0}^\infty\frac{(z+\lambda)^n}{\sqrt{n!}}\;\fn
        = G(z)e^{\lambda (\Re (z) + \frac {\lambda}2 )}\eta_{z + \lambda}\; .
\end{equation}

Comparing (\ref{phot-add-bin3}) with (\ref{transf-nlCS}) and
writing $\eta^{F}_{z}=\zeta^{}_{z+\la}$, it can be  found that
$w=z+\la$, $x^{}_{n}=n$ and $\psi^{}_{n}=\fn$. Furthermore,
$\N^\prime (\vert w \vert^2 ) = e^{\vert z \vert^2} \vert
G(z)\vert^{-2}$ and $\Omega (w ) = e^{i\Theta (w )}$, where It is
written $G(z ) = \vert G(z)\vert e^{i\Theta (w )}$.   It is
remarkable that in this example while $\eta^{F}_{z}$ is written in
(\ref{phot-add-CS}) in terms of a non-orthonormal basis
$\{\phi_n^F\}_{n=0}^\infty$, when these vectors are considered as
constituting a basis for $\HH$, its transcription in terms of the
orthonormal basis $\{\phi_n\}_{n=0}^\infty$ only involves a shift
in the variable $z$ and no change in the components.

It is now straightforward to write down a resolution of identity,
following the pattern of the canonical CSs. Indeed, writing
$w=z+\la=\rho e^{i\theta}$, we have (on $\HH$),
\begin{equation}\label{resolid2}
\int\!\!\int_{\C}|\zeta_{w}\rangle \langle
\zeta_{w}|\N'(|w|^{2})d\mu(w,\overline{w})=\hat I \; , \qquad
d\mu(w,\overline{w})=\frac{e^{-\rho^{2}}}{\pi}\rho d\rho
d\theta\; .
\end{equation}

The dual CSs $\eta^{F^{-1}}_{z}$ are obtained by replacing the
$\phi^{F}_{n}$ in (\ref{phot-add-bin}) by $\phi^{F^{-1}}_{n}=\hat
T\fn=G(a)^{-1}e^{-\la a^\dag} \fn$. But since $G(a)^{-1}e^{-\la
a^\dag}=e^{-\la a^\dag }G(a- \la \hat I)^{-1} =e^{-\la a^
\dag}G(a-\la \hat I)^{-1}\fn.$ Hence, using the same argument as
with the $\phi^F_{n}$, we will observe that in the present case
(up to normalization), the dual pair of states $\eta^F_z$ and
$\eta^{F^{-1}}_z$ is obtained simply by replacing $\la$ by $-\la$.

Two particular cases of the operator  $\hat T^{-1}$ in
(\ref{ex1T}) are of special interest. In the first instance take
$G(a)=\hat I$, so that $\hat T^{-1}=e^{\la a^\dag}$. The vectors
$\phi^{F}_{n}=\hat T^{-1}\fn$ may easily be calculated. Indeed we
get
\begin {equation}\label{phot-add1}
\phi^{F}_{n}=\sum_{k=0}^{\infty}\frac{(\la a^\dag)^{k}}
{k!}\fn=e^{\frac{\la^{2}}{2}}\frac{{a^\dag}^{n}}{\sqrt{n!}}\eta^{}_{\la}\;
,
 \end {equation}
which (up to  normalization) are the well-known {\em photon-added
CSs\/} of quantum optics \cite{Agarval-tara, roymeh}. Hence in
this case we write $\phi^{F}_{n}=\phi^{\rm{pa}}_{\la, n}$. We
denote the corresponding CSs by $\eta^{\rm{pa}}_{\la, z}$ and
note that
\begin{equation}\label{phot-add2}
\eta^{F}_{z}:=\eta^{\rm{pa}}_{\la,z} =
\N(|z|^{2})^{-\frac{1}{2}}\sum_{n=0}^{\infty}\frac{z^{n}}{\sqrt{n!}}\phi^{\rm{pa}}_{\la
,n}= e^{\la(x+\frac{\la}{2})}\eta^{}_{z+\la}\; ,
\end{equation}
where $x=\Re(z)$. Clearly if $\la\longrightarrow 0$, then
$\eta^{\rm{pa}}_{\la,z}\longrightarrow\eta^{}_{z}$. It ought to
be emphasized at this point, however, that while the vectors
$\phi^{\rm{pa}}_{\la, n}$ are (up to normalization) photon added
CSs, the vectors $\eta^{\rm{pa}}_{\la, z}$ are just (up to
normalization) canonical CSs. The dual set of CSs,
$\eta^{F^{-1}}_{z}$ are obtained by replacing $\la$ by $-\la$ so
that the states $\eta^{\rm{pa}}_{\la, z}$ and
$\eta^{\rm{pa}}_{-\la, z},\; z \in \mathbb C$,  are in duality,
and the interesting relation,$ \langle\eta^{\rm{pa}}_{-\la,
z}|\eta^{\rm{pa}}_{\la, z} \rangle_{\HH}=e^{-\la(\la+2iy)}$ holds.

On $\HH^{}_{F}$ one has the creation and annihilation operators
(see (\ref{ex1-adj3})),
\begin {equation}\label{phot-add4}
a^{}_{F}=a-\la \hat I,\quad a^\dag_{F}=a^\dag\; ,
\end {equation}
which are adjoints of each other on $\HH^{}_{F}$, but clearly not
so on $\HH$. However, on $\HH$  the two operators $A$ and
$A^\dag$ are as in (\ref{ex1-adj4}):
$$ A=a-\la \hat I,\quad A^\dag=a^\dag-\la \hat I\; . $$

   As the second particular case of (\ref{ex1T}), we
take $\la=0$ and $G(a)=e^{\mu a},\;  \mu\in\Re,$ i.e., $\hat
T^{-1}=e^{\mu a}$. The basis vectors are now
\begin {equation}\label{bin1}
\phi^{F}_{n}=e^{\mu
a}\fn=\sqrt{n!}\sum_{k=0}^{n}\frac{\mu^{n-k}}{\sqrt{k!}(n-k)!} \;
\phi_k=\frac{(a^\dag+\mu \hat I)^{n}}{\sqrt{n!}}\;\phi_{0}.
\end{equation}
These states have also been studied in the quantum optical
literature \cite{Fufengsol} and in view of the last expression in
(\ref{bin1}), we shall call them {\em binomial states} and write
$\phi^F_n=\phi^{\rm{bin}}_{\mu , n}$. The CSs, built out of these
vectors as basis states, are:
\begin{eqnarray}\label{bin2}
\eta^F_z:=\eta^{\rm{bin}}_{\mu,z} &=& e^{\mu
a}\eta_z=e^{-|z|^2/2}\sum_{n=0}^\infty\frac{z^n}{\sqrt{n!}}\; \phi^{\rm{bin}}_{\mu, n}\nonumber\\
&=& e^{\mu x-|z|^2/2}\sum_{n=0}^\infty\frac{z^n}{\sqrt{n!}}\;\fn.
\end{eqnarray}
The dual family of CSs are simply $\eta^{\rm{bin}}_{-\mu,z}$ and
$ \langle\eta^{\rm{bin}}_{-\mu,z}|\eta^{\rm{bin}}_{\mu,z}
\rangle=1$. The creation and annihilation operators on $\HH_F$
are:
\begin{equation}\label{bin4}
  \af=a\; ,\qquad \afd=a^\dag+\la \hat I\; ,
\end{equation}
while the other two operators on $\HH$ are:
\begin{equation}\label{bin5}
  A=a\; ,\qquad A^\dag=a^\dag\; .
\end{equation}
The operators (\ref{bin4}) have been studied, in the context of
on-self-adjoint Hamiltonians in \cite{Beckera, bedeszaf,
Becker2001}. Again, it is remarkable that the CSs
$\eta^{\rm{bin}}_{\mu,z}$ are exactly the canonical CSs, $\eta^{
}_z$, up to a factor.

 At the  end of this subsection, I briefly remind that we have
 established in Ref. \cite{rokntvs} the relation
\begin {eqnarray}\label{bin1}
 \phi_{m,\mu}&=&\frac { e^{\mu a} }    { \sqrt
 {L_m^{(0)}(-\mu^2)} }\; \phi_m
 =\sum_{n=0}^{m}\frac{ \sqrt{m!}\;
 \mu^{m-n}}{(m-n)!\sqrt{n! L_m^{(0)}(-\mu^2) }} \; \phi_n
 \nonumber \\ &=& \left(\frac{{\mu}^{2m }m!}{L_m^{(0)}(-\mu^2)}
 \right)^{1/2} \sum _{n=0}^m \frac {(\mu^{-1})^n} {\sqrt
 {n!}[(m-n)]!} \; \phi_n = |\mu^{-1},f \rangle,
\end{eqnarray}
where $L_m^{(0)}(-\mu^2)$ are the $m-$order Laguerre functions
and the states $|\mu^{-1},f \rangle$ have been demonstrated as the
bound state nonlinear CSs. Then we have used the states in
(\ref{bin1}) as the basis of a $f-$deformed Fock space (with
nonlinearity function $f(\hat n) = (m- \hat n)$) and finally the
representations of coherent and squeezed states in these new
basis together with their statistics have been studied in detail
\cite{rokntvs}.
\subsubsection{Re-scaled basis states and nonlinear CSs}
For the next general class of examples, let the operator $\hat
T^{-1}$ have the form
\begin{equation}\label{ex2T}
 \hat  T^{-1}:= \hat T(\hat n)^{-1}=\sum_{n=0}^\infty
 \frac{1}{t(n)}|\fn\rangle\langle\fn|,
\end{equation}
where  $t(n)$s are real numbers, having the properties:
\begin{enumerate}
  \item $t(0)=1$ and $t(n)=t(n')$ if and only if $n=n'$\; ;
  \item $0<t(n)<\infty$\; ;
  \item the finiteness condition for the limit
\begin{equation}\label{Tfin}
 \lim_{n\to\infty}\left[\frac{t(n)}{t(n+1)}\right]^2\cdot\frac{1}{n+1}=\rho<\infty
\end{equation}
holds, which implies that the series
\end{enumerate}
 \begin{equation}\label{Tfin2}
  \sum_{n=0}^\infty\frac{r^{2n}}{[t(n)]^2n!}:=S(r^2),
\end{equation}
converges for all $r<L=1/\sqrt{\rho}$. The operators $\hat T$ and
$\hat F$ are now
\begin{equation}\label{ex2TF}
 \hat T := \hat T(\hat n)  =  \sum_{n=0}^\infty t(n)|\fn\rangle\langle\fn|\; , \qquad
 \hat F := \hat F(\hat  n) = \sum_{n=0}^\infty t(n)^2|\fn\rangle\langle\fn|\; .
\end{equation}
Let me define a new operator $f(\hat n)$,  by its action on the
basis vectors.
\begin{equation}\label{nonlinf}
  f(\hat n)\fn :=\frac{t(n)}{t(n-1)}\fn =f(n)\fn,
\end{equation}
then
\begin{equation}\label{nonlinf2}
  t(n)=f(n)f(n-1)\cdots f(1):=[f(n)]!.
\end{equation}
Thus one has the transformed, non-orthogonal basis vectors
\begin{equation}\label{nonlinf3}
  \phi^F_n=\frac{1}{t(n)}\fn=\frac{1}{[f(n)]!}\fn \; ,
\end{equation}

We shall call the vectors (\ref{nonlinf3}) {\em re-scaled basis
states\/}.

      The CSs $\eta^F_z$ are now:
\begin{equation}\label{nonlinCS}
  \eta^F_z=\N(|z|^2)^{-1/2}\sum_{n=0}^\infty\frac{z^n}{\sqrt{n!}}\phi^F_n,
\end{equation}
which, as vectors in $\HH_F$ are well defined and normalized for
all $z\in \C$. However, when considered as vectors in $\HH$ and
rewritten as:
\begin{equation}\label{nonlinCS2}
  \eta^F_z=\N(|z|^2)^{-1/2}\sum_{n=0}^\infty\frac{z^n}{[f(n)]!\sqrt{n!}}\fn,
\end{equation}
are no longer normalized and defined only on the domain (see
(\ref{Tfin}) and (\ref{Tfin2})),
\begin{equation}\label{nonlindom}
  \D=\left\{z\in\C\Big| \;|z|<L=\frac{1}{\rho}\right\}.
\end{equation}
The operators $\af$ and $\afd$ act on the vectors $\phi^F_n$
 as
\begin{equation}\label{nonlinops}
  a_F\phi^F_n=\sqrt{n}\phi^F_{n-1}\; ,\qquad
  \afd\phi^F_n = \sqrt{n+1}\phi_{n+1}^F\; .
\end{equation}
The operator $A=a_F$, considered as an operator on $\HH$ and its
adjoint $A^\dag$ on $\HH$ act on the original basis vectors $\fn$
in the manner,
\begin{equation}\label{nonlinops2}
  A\fn =  f(n)\sqrt{n}\phi_{n-1}\; , \qquad
  A^\dag\fn  =  f(n+1)\sqrt{n+1}\phi_{n+1}\; ,
\end{equation}
and thus, they can be  written in an obvious notation as
\begin{equation}\label{nonlinops3}
  A=af(\hat n)\; ,\qquad A^\dag=f(\hat n)a^\dag\; ,
\end{equation}
as operators on $\HH$.

Thus, up to a normalization factor, the CSs defined in
(\ref{nonlinCS2}) are the well-known {\it nonlinear CSs} of
quantum optics \cite{Manko1997, Matos1996}.

 The dual CSs $\eta^{F^{-1}}_z$
 which, as vectors in the Hilbert space $\HH_{F^{-1}}$,
 will be well-defined vectors in $\HH$ only if
\begin{equation}\label{nonlindualdom}
  \lim_{n\to\infty}\left[ \frac{t(n+1)}{t(n)}\right]^2\cdot\frac{1}{n+1}=\widetilde{\rho}<\infty.
\end{equation}
From now on, the sign "tilde" over the "operators" and "states",
assign them to the corresponding "dual operators" and "dual
states", respectively. In this case, one has
\begin{equation}\label{nonlinCSdual}
  \eta^{F^{-1}}_z =
  \widetilde{\eta}\;^{F}_z=\N(|z|^2)^{-1/2}\sum_{n=0}^\infty\frac{[f(n)]!z^n}{\sqrt{n!}}\;\fn,
\end{equation}
and are defined (as vectors in $\HH$) on the domain
\begin{equation}\label{nonlindualdom2}
\widetilde{\D}=\left\{z\in\C\Big|
|z|<\widetilde{L}=\frac{1}{\sqrt{\widetilde{\rho}}}\right\}.
\end{equation}
 Equations (\ref{nonlinCSdual}) and (\ref{nonlindualdom2}) should
 be compared to (\ref{nonlinCS2}) and (\ref{nonlindom}). One also
 has
  $\langle\eta_z^{F^{-1}}\Big |\eta_z^F\rangle_\HH=1,$
  for all $z\in \D\cap\widetilde{\D}$.

A resolution of the identity  on $\HH$ can be obtained in terms of
the vectors $\eta_z^F$ (or $\eta_z^{F^{-1}}$) by solving the
moment condition
\begin{equation}\label{nonlinmomcond}
  \int_0^L r^{2n}d\la(r)=\frac{\left[f(n)! \right]^2
  n!}{2\pi},\quad n=0,1,2,\cdots \; .
\end{equation}

  A highly instructive example of the duality between families of nonlinear CSs
   is provided by the GP CSs $|z, \kappa\rangle_{GP}=
\eta^{\rm{GP}}_z$ introduced in (\ref{gilperCS}) \cite{gilmore}
and BG CSs $|z, \kappa\rangle_{BG}= \eta^{\rm{BG}}_z$ introduced
in (\ref{BG1}) \cite{bar-gir71}, defined for the discrete series
representations of the group $SU(1,1)$.  It is now immediately
clear that the operator
\begin{equation}\label{bargir-gilper-trans}
  \hat T(\hat  n)=\sum_{n=0}\frac{1}{\sqrt{(2\kappa
  +n-1)!}}|\fn\rangle\langle\;\fn|\;,
\end{equation}
acts in the manner,
\begin{equation}
  \eta_z^{\rm{BG}} = \lambda_1 \; \hat T(\hat n)\eta_z\; \qquad \rm{and} \qquad
  \eta_z^{\rm{GP}} = \lambda_2\; \hat T(\hat n)^{-1}\eta_z\; ,
\label{bargir-gilper-trans2}
\end{equation}
where $\lambda_1$ and $\lambda_2$ are constants, thus
demonstrating the relation of duality between the two sets of CSs.
  \section{ Introducing the generalized displacement operators and the dual family of MP CSs}
  \label{sec-displce}
   After we recasted  the KPS, HG, PS, TC and $su(1, 1)$  CSs
   as nonlinear CSs and found a nonlinearity
   function for each set of them, it is now possible to construct all of
   them through a displacement type operator
   formalism. I will do this in two distinct ways.\\
 {\bf I)} Applying the mathematical physics formalism for each
   set of the nonlinear CSs have been established in section 3
   which obtained by the action of $\hat T^{-1}(f(\hat n))$ on canonical CSs, we
   are now ready to introduce the dual family associated with each class of them, using the action of
   $\hat T (f(\hat n))$ on the canonical CSs. But this is not the only
   formalism to find the dual family. In what follows
   another formalism presented by B Roy and P Roy will be explained \cite{royroy}.\\
 {\bf II)}  According to the proposition in \cite{royroy}
            the authors have been defined two new operators:
 \begin{equation}\label{kps-25}
   B=a\frac{1} {f(\hat{n})},  \qquad
   B^\dag=\frac{1}{f(\hat{n})}a^\dag,
 \end{equation}
    such that the canonical commutation relations $[A,B^\dag]=1=[B,
    A^\dag]$ hold.
    Before we proceed any further, to clarify more the problem,
    an interesting result may be given here.
    Choosing  special compositions of the operators $A$, $A^\dagger$  in (\ref{nonl-annih})
    and $B$, $B^{\dag}$ in (\ref {kps-25}), we may observe that $B^{\dag}A|n \rangle=n|n \rangle
    =A^{\dag}B|n \rangle$.
     It can easily be investigated that the generators $\{A, B^{\dag}, B^{\dag}A, \hat I\}$ constitute the
    commutation relations of the Lie algebra $h_4$.
    The corresponding Lie group is the well-known
    Weyl-Heisenberg(W-H) group denoted by $H_4$.
    The same situation holds for the set of generators $\{B, A^{\dag}, A^{\dag}B,
    \hat I\}$.

    Coming back again to the Roy and Roy formalism,
    the relations in (\ref{kps-25}) allow one to define two generalized displacement operators
 \begin{equation}\label{kps-26}
   \widetilde D_f (z)=\exp(zA^\dag-z^ *B), \qquad  \widetilde D_f
   (z)|0\rangle = |\widetilde{z,f}\rangle,
 \end{equation}
 \begin{equation}\label{kps-27}
   D_f(z)=\exp(zB^\dag-z^* A), \qquad   D_f
   (z)|0\rangle = |{z,f}\rangle.
 \end{equation}
     Noting that $\widetilde D_f(z)=D_f(-z)^\dag = [D_f(z)^{-1}]^\dag$,
     it may be realized that the {\it dual} pairs obtained generally from the actions of
     (\ref {kps-26}) and (\ref {kps-27}) on the vacuum state are
     the orbits of a projective {\it nonunitary} representations of the
     W-H group \cite{AliRokTav}, so we named {\it displacement type} or {\it generalized
     displacement} operator.
   Therefore (using each one of the above two formalisms)
   it is possible to construct a new class of CSs related to each set of
   MP CSs, $|z, f\rangle_{\rm Mp}$ introduced in section 3.
 \vspace{4 mm}\\
   {\it Example $1\;$ The dual family of KPS CSs:}
   \vspace{2 mm}\\
   Considering  the
   KPS  introduced in \cite{kps}, one can see that:
 \begin{equation}\label{kps-28}
     D_f(z)|0\rangle \equiv |z\rangle_{_{\rm KPS}}
 \end{equation}
   and the other one which is a new family of CSs, named {\it
   "dual states"} in \cite{AliRokTav, royroy} is:
 \begin{equation}\label{kps-29}
   \widetilde D_f(z)|0\rangle \equiv |\widetilde z\rangle_{_{\rm KPS}}
   = \widetilde{\N} (|z|^2)^{-1/2}\sum_{n=0}^{\infty}\frac{z^n
   \sqrt{\rho(n)}} {n!}|n\rangle,
 \end{equation}
   where the normalization constant is determined as:
   $ \tilde {\N} (|z|^2) =
  \sum_{n=0}^{\infty}\frac{|z|^{2n}\rho(n)}{(n!)^2}.$
   Obviously the states $|\tilde z\rangle_{_{KPS}}$ in (\ref {kps-29})
   are new ones, other than
   $|z\rangle_{_{KPS}}$. By the same procedures we have done in the
   previous sections it may be seen that the new states,
   $|\tilde z\rangle_{_{KPS}}$ also can be considered as nonlinear CSs with the nonlinearity function
 \begin{equation}\label{kps-60}
   \tilde f_{_{KPS}}(\hat{n})=\sqrt {\frac{\hat{n}
   \rho(\hat{n}-1)}{\rho(\hat{n})}},
 \end{equation}
   which is exactly the inverse of $f_{_{\rm KPS}}(\hat{n})$ derived in (\ref{kps-20}), as one may
   expect. Also the Hamiltonian for the dual oscillator is found
   to be
 \begin{equation}\label{dual-H}
   \tilde{\hat{H}}_{_{\rm KPS}}
   =\hat{n}\left(\tilde f_{_{\rm KPS}}(\hat{n})\right)^2=\hat{n}^2
   \hat{H}_{_{\rm KPS}}^{-1}.
 \end{equation}

    The above results may also be obtained using the mathematical physics
   formalism.
   One can define the operator
 \begin{equation}\label{Top}
   \hat{T}=\sum_{n=0}^{\infty}\sqrt{\frac{n!}{\rho(n)}}|n \rangle \langle
   n|,
 \end{equation}
   the action of which on canonical CSs, $|z\rangle_{_{CCS}}
   =\exp{(-|z|^2/2)}\sum_{n=0}^{\infty}\frac{z^n}{\sqrt
   n!}|n\rangle$, yields the KPS CSs:
 \begin{equation}\label{Top1}
   \hat{T} |z\rangle_{_{\rm CCS}} = |z\rangle_{_{\rm KPS}}.
 \end{equation}
   The $\hat{T}$ operator we introduced in Eq. (\ref {Top}) is well-defined and
   the inverse of it can be easily obtained as
 \begin{equation}\label{T}
   \hat{T}^{-1} = \sum_{n=0}^{\infty}\sqrt{\frac{\rho(n)}{n!}}|n \rangle \langle
   n|,
 \end{equation}
   by which we may construct the new family of dual states:
   $\hat{T}^{-1} |z\rangle_{_{\rm CCS}}  \equiv |\tilde z\rangle_{_{\rm
   KPS}}$, which are just
   the states have been obtained in (\ref{kps-29}). So, in what
   follows each of the two approaches which are more easier have been
   used for constructing the dual pair of each MP CSs.

   Applying the presented formalism on the numerous weight
   functions $\rho(n)$ of the KPS type in Ref. \cite{kps} is now easy, so our intention is not to refer to them
   explicitly (it may be found in \cite {Roknizadeh2004}, completely).
   Our reconstruction of these states by the standard definitions,
   i.e. annihilation operator eigen-states and displacement operator
   techniques,  enriches each set of
   the above classes of CSs in quantum
   optics, in the context of each other.
 \vspace{4 mm}\\
   {\it Example $2\;$ The dual family of HG  CSs:}
   \vspace{2 mm}\\

  As another class of generalized CSs, we observed that the HG  CSs
  in (\ref {GHCS}) can be constructed by starting with the hypergeometric function.
  Now, going back to these states  we apply to    the canonical CSs on
  $\HH$, the operators
\begin{eqnarray}
 \hat T  &:=&   \sum_{n=0}^\infty \left[\frac {(\alpha_1)_n \ldots (\alpha_p)_n}
           {(\beta_1)_n \ldots (\beta_q)_n}\right]^{\frac 12}\;\vert\fn\rangle\langle\fn\vert\; ,
           \nonumber\\
 \hat T^{-1}  &:=& \sum_{n=0}^\infty \left[\frac {(\alpha_1)_n  \ldots (\alpha_p)_n}
           {(\beta_1)_n  \ldots (\beta_q)_n}\right]^{-\frac
           12}\;\vert\fn\rangle\langle\fn\vert\;.
            \end{eqnarray}
The explicit form of the dual states as a  new set of again HG
CSs are as follows (not of the HG  CSs type of \cite{Appschill}):
\begin{equation}\label{DGHCS}
| \widetilde{z; p, q} \rangle = | \alpha_1, \cdots \alpha_p;
\beta_1,
 \cdots \beta_q; z \rangle = {}_p \widetilde{\N}_q (|z|^2) ^{-1/2}
\sum_{n=0}^\infty \frac{z^n \sqrt {_p \rho _q(n)}}{n!}\;|n\rangle
\end{equation}
where $_p \rho _q(n)$ are defined by the relation (\ref {p-rho-q})
and the normalization factor is determined as
\begin{equation}
 {}_p \widetilde{\N}_q (|z|^2) =
 {}_pF_q (   \beta_1, \ldots , \beta_q \; ; \alpha_1, \ldots , \alpha_p ;   \; x ).
\end{equation}
It is then immediate that the corresponding families of CSs
$\{\eta_z^F\}$ and $\{\eta_z^{F^{-1}}\}$ will be in duality.
(Actually, it may be necessary to impose additional restrictions
on the $\alpha_i$ and $\beta_i$, in order to ensure that the CSs
$\{\eta_z^F\}$ and $\{\eta_z^{F^{-1}}\}$, when defined on $\HH$,
satisfy a resolution of the identity \cite{Appschill}).
 \vspace{4 mm}\\
   {\it Example $3\;$ The dual family of PS CSs:}
   \vspace{2 mm}\\
   The dual of the PS states (\ref {ps1}) can be easily obtained using
   the approach in \cite{royroy}:
 \begin{equation}\label{ps5}
   |\widetilde{q, z}\rangle_{\rm {PS}} = \widetilde  \N (q, |z|^2)^{-1/2} \sum_{n=0}^{\infty}
   \frac{q^{-n(n-1)/2}}{\sqrt{n!}}z^n |n\rangle,
 \end{equation}
   where $\widetilde \N (q, |z|^2)$ is some  normalization constant, may be
   determined. For this example the proposition in
   \cite{AliRokTav}
   works well. The $\hat{T}$-operator in this case reads:
 \begin{equation}\label{ps6}
   \hat{T}= \sum_{n=0}^{\infty}
   q^{\hat n(\hat n-1)/2} |n\rangle \langle n|,
 \end{equation}
   by which one may obtain:
 \begin{equation}\label{ps7}
   \hat{T}^{-1}|z\rangle_{\rm {CCS}} \equiv \left(\sum_{n=0}^{\infty}
   q^{-n(n-1)/2} |n\rangle \langle n|\right)  |z\rangle_{\rm CCS}=|\widetilde{z, q}\rangle_{_{\rm
   PS}}.
 \end{equation}
    which is exactly the dual states we obtained in
    Eq. (\ref{ps5}).
   \vspace{4 mm}\\
 {\it Example $4\;$ The dual family of TC CSs}
   \vspace{2 mm}\\
   The dual family of TC CSs of the first kind, may be obtained by either of the two
   formalisms. Anyway, the results is as follows:
 \begin{equation}
   |\widetilde{z;p}\rangle_{\rm TC}^{(1)} = \widetilde {\N}_p
   (|z|^2)^{-1/2} \sum_{n=0}^\infty \frac{z^n} {\sqrt{n!}}
   \; \sqrt{d_p(n)}\; |n\rangle.
 \end{equation}
And similarly for the second kind one has:
\begin{equation}
|\widetilde{z; \lambda, \beta}\rangle_{\rm TC}^{(2)} = \widetilde
{\N}_{\lambda, \beta} (|z|^2)^{-1/2}\sum_{n=0}^{\infty}
\frac{z^n} {\sqrt{n!}} \; \sqrt{d_{\lambda, \beta }(n)}
 \;   |n\rangle.
\end{equation}
where $\widetilde{\N}_p (|z|^2) = \sum_{n=0}^\infty
\frac{|z|^{2n}d_p(n)}{n!}$ and $ \widetilde {\N}_{\lambda, \beta}
(|z|^2) = \sum_{n=0}^\infty \frac{|z|^{2n}d_{\lambda,
\beta}(n)}{n!}$. The parameters used in these relations are the
same as the ones have
been explained after Eq. (\ref{TC-2}).\\
   {\it Example $5\;$ Generalized displacement operators for BG and GP
   CSs of $su(1, 1)$ Lie algebra:}
   \vspace{2 mm}\\
   As a well-known example we express the dual of
   BG CSs (of $SU(1, 1)$) group.
   The duality of these states with the so-called
   GP CSs (of $SU(1, 1)$) have already been demonstrated in \cite{AliRokTav}.
   The latter states were defined as:
 \begin{equation}\label{GP1}
   |z,\kappa\rangle_{_{\rm GP}}=\N(|z|^2)^{-1/2}\sum_{n=0}^{\infty}\sqrt{\frac{(n+2\kappa-1)!}{n!}}
   z^n |n\rangle.
 \end{equation}
   where $\N (|z|^2)$ is a normalization constant.
   The nonlinearity function and the Hamiltonian may be written
   as $f_{_{\rm GP}}(\hat{n})=f_{_{\rm BG}}^{-1}(\hat{n})$ and
   $\hat{H}=\hat{n}/(\hat{n}+2\kappa-1)$, respectively. Therefore we have in
   this case:
 \begin{equation}\label{GP2}
   B|\kappa, n\rangle=\sqrt{\frac{n}{n+2\kappa-1}}|\kappa,
   n-1\rangle,
 \end{equation}
 \begin{equation}\label{GP3}
   B^{\dag}|\kappa, n\rangle=\sqrt{\frac{n+1}{n+2\kappa}}|\kappa,
   n+1\rangle,
 \end{equation}
 \begin{equation}\label{GP4}
   [B, B^{\dag}]|\kappa, n\rangle=\frac{2\kappa-1}{(n+2\kappa)(n+2\kappa-1)}|\kappa,
   n\rangle.
 \end{equation}
   where $B=af_{_{GP}}(\hat{n})$ and $B^{\dag}=f_{_{GP}}(\hat{n})a^{\dag}$
   are the deformed annihilation and creation
   operators for the dual system (GP CSs), respectively. It is
   immediately observed that $[A, B^{\dag}]=\hat I, [B, A^{\dag}]=\hat I
   $, where obviously $A=af_{_{BG}}(\hat{n})$ and $A^{\dag}$ is its
   Hermition conjugate. So it is possible to obtain the
   displacement type operators for the BG and GP
   (nonlinear) CSs discussed in this tutorial, using relations (\ref{kps-26}) and
   (\ref{kps-27}). As a result the displacement operators
   obtained by presented method are such that:
 \begin{equation}\label{GP6}
   |z,\kappa\rangle_{\rm {BG}}=D_{\rm {BG}}(z)|\kappa, 0\rangle=
   \exp{(zA^{\dag}-z^*B)}|\kappa, 0\rangle,
 \end{equation}
    and
 \begin{equation}\label{GP7}
   |z,\kappa\rangle_{\rm {GP}}=D_{\rm {GP}}(z)|\kappa,
    0\rangle=\exp{(zB^{\dag}-z^*A)}|\kappa, 0\rangle.
 \end{equation}

   To apply the procedure to the $su(1, 1)$-BG
   CSs for Landau levels we expressed as another example, one must re-defined the auxiliary operators $B$ and
   $B^\dag$, in place of the ones introduced in (\ref {kps-25}), as follows:
\begin{equation}\label{BB-LL}
   B=\frac{1}{f_{\rm {LL}}(\hat{n})}a, \qquad B^\dag =a^\dag
   \frac{1}{f_{\rm {LL}}(\hat{n})}.
\end{equation}
\section{The link between GK-CSs and nonlinear CSs}
    As it has mentioned earlier the GK-CSs  can not
    be fully placed in the above two formalisms.  We will pay attention
    to this matter in the following section. Indeed we have to try
    some {\it radically different method} from the previous ones.
\subsection{A discussion on the modification of GK-CSs}\label{sec-Discusson}
    As it is observed  previously in subsection 2.6.1, in the
    modification imposed by El Kinani and Daoud on the GK-CSs, the parameter $\alpha$
    has been {\it implicitly} considered as a constant,
    which its presence in the exponential factor
    of the introduced CSs preserves the temporal stability requirement
    (it is not now an integration {\it variable}).
    Meanwhile, for the
    temporal stability of the GKCSs in (\ref {GKED}) one reads:
 \begin{equation}\label{TS}
    e^{-i\hat{H}t}|z, \alpha\rangle = |z, \alpha'\rangle, \qquad
    \alpha'=\alpha + \omega t.
 \end{equation}
    Upon a closer inspection, one can see that the latter relation is indeed
    inconsistent with the
    resolution of the identity. By this we mean that when $\alpha$
    is considered as a constant parameter, it really labels any  over-complete set of GKCSs,
    $\{ |z, \alpha\rangle \}$. But the time evolution
    operator in (\ref{TS}) maps the over-complete set of states $\{ |z, \alpha\rangle \}$
    to another over-complete set $\{ |z, \alpha'\rangle \}$.
    These are two {\it distinct set of CSs}, each labels with a
    specific $\alpha$, if one consider the El kinani-Daoud formalism.
    But the temporal stability precisely means that the time evolution
    of a CS remains a CS, {\it of the same
    family}. So transparently speaking, the states
    introduced in (\ref{GKED}) are not of the Gazeau-Klauder type, exactly.

    To overcome this problem,
    we redefine the resolution of the identity as follows:
 \begin{equation}\label{RI-RT}
  \lim_{\Gamma\rightarrow
   \infty}\frac{1}{2\Gamma}\int_{-\Gamma}^{\Gamma} d\alpha \int_0^R
    |z, \alpha\rangle \langle z,
     \alpha|d\lambda(z)=\sum_{n=0}^\infty|n\rangle\langle n|=
     \hat{I},
      \qquad 0 < R \leq \infty.
 \end{equation}
     One can simplify the LHS of
    (\ref{RI-RT}) which interestingly led exactly to the LHS of
    (\ref{RI-GKED}). Indeed one gets:
\begin{equation}\label{}
  \int_0^{R}|z, \alpha\rangle \langle z, \alpha|d \lambda(z)=
   \lim_{\Gamma\rightarrow
     \infty}\frac{1}{2\Gamma}\int_{-\Gamma}^{\Gamma} d\alpha \int_0^R
      |z, \alpha\rangle \langle z, \alpha|d\lambda(z),
 \end{equation}
   where $d\lambda(z)$ is determined as in (\ref{measure}).

  By this fact it may be concluded that both of
  the over-complete collection of states:
  $\{|z, \alpha\rangle \}$ and $\{|z, \alpha'\rangle\}$, when $\alpha$ and
  $\alpha'\equiv\alpha +\omega t$ both are {\it fixed},
  belong to a large set of over-complete states
  with an arbitrary $\alpha$:
\begin{equation}\label{largeH}
  \{|z, \alpha\rangle, z\in C, -\infty \leq \alpha \leq \infty\}.
\end{equation}
  Note that by replacing $\alpha \in R$ with $-\infty \leq \alpha \leq
  \infty$ in (\ref{largeH}) we want to emphasis that we relax $\alpha$
  from the constraint of being fixed.
  But unfortunately, the {\it variability} of $\alpha$ destroys the well definition of
  the operator $f(\alpha, \hat{n})$ will be
  introduced later in (\ref {nlGKED})
  and therefore deformed annihilation and creation operators $A$ and $A^\dag$.
  To overcome this difficulty we may bridge the gap between these two situations:
  variability and constancy of $\alpha$.
  We define the set of operators $A=a f(\alpha, \hat n)$, $A^\dag=f^\dag(\alpha, \hat n) a^\dag$
  and any other
  operator which explicitly depends on $\alpha$,
  in each {\it sector} ({\it subspace}) $\HH_\alpha$, labeled by a specific $\alpha$
  parameter, of the whole Hilbert space $\HH$
  which contains all GKCSs $\{|z, \alpha \rangle\}$.
  Indeed the whole Hilbert space foliates by each
  $\alpha$ (remember the continuity of $\alpha$).
   Moreover the action of the time evolution
   operator on any state on a specific sector, falls it down to another
   sector, both belong to a large Hilbert space.
   So, {\it when one deals with the operators that depend on the $\alpha$ parameter, it  should necessarily be fixed,
   while this is not the case when we are dealing with the states.}

\subsection{The relation between nonlinear CSs and GKCSs}\label{sec-GKCS-NL}
   Following the second formalism presented in this article  for the states expressed
   in (\ref {GKED}) one may obtain \cite{Roknizadeh2004}
  \begin{eqnarray}\label{nlGKED}
    f_{\rm {GK}}(\alpha, \hat{n}) &=& e^{i\alpha(\hat{e}_n-\hat{e}_{n-1})}
    \sqrt{\frac{\rho(\hat{n})}{\hat{n}\rho(\hat{n}-1)}},
    \qquad \alpha \; \rm{being \; fixed}\nonumber\\
    &=& e^{i\alpha(\hat{e}_n-\hat{e}_{n-1})} \sqrt {\frac {\hat e_n}{\hat
    n}},
  \end{eqnarray}
    where the notation $\hat{e}_n \equiv
    \rho(\hat{n})/\rho(\hat{n}-1)$ has been choosed.

   Moreover, we gain the opportunity to find rising and lowering operators
   in a safe manner:
\begin{equation}\label{crea-anni}
  A_{\rm GK}=a f_{\rm GK}(\alpha, \hat{n}),  \qquad A^\dag_{\rm GK}=
  f^\dag_{\rm GK}(\alpha, \hat{n})a^\dag,
\end{equation}
    where one may easily verify that $A_{\rm GK}|z, \alpha \rangle = z |z,
    \alpha\rangle$.
    Obviously the commutation relation between these two
   ($f-$deformed) ladder operators obeys the relation \cite{Manko1997}:
\begin{eqnarray}\label{com}
  [A_{\rm GK}, A^\dag_{\rm GK}] & = &\frac {\rho(\hat{n}+1)}{\rho(\hat{n})} -
  \frac {\rho(\hat{n})}{\rho(\hat{n}-1)}\nonumber\\
  &=& \hat{e}_{n+1}-\hat{e}_n.
\end{eqnarray}
   The special case $\rho(n)=n!$ recovers the standard bosonic
   commutation relation $[a, a^\dag]=\hat{I}$.
   Using the {\it "normal-ordered"} form of the Hamiltonian as in
    \cite{Roknizadeh2004} and taking $\hbar=1=\omega$,
     for the Hamiltonian of GKCSs we get
  \begin{equation}\label{normalH}
   \hat{H}_{\rm GK}\equiv \HD = A_{\rm GK}^\dag A_{\rm GK} = \hat{n}
   \Big|f_{\rm GK}(\alpha, \hat{n})\Big|^2 =
   \frac{\rho(\hat{n})}{\rho(\hat{n}-1)}=\hat{e_n}
  \end{equation}
   shows clearly the independency of the dynamics of the system of $\alpha$.
   By the above explanations
    the deformed annihilation and creation operators $A_{\rm GK}$
    and $A^\dag_{\rm GK}$ of the oscillator algebra, satisfy the eigenvector equations:
 \begin{equation}\label{GK-an1}
   A_{\rm GK}|n\rangle = \sqrt{ e_n}
   e^{i\alpha ( e_n-e_{n-1})}|n-1\rangle,
 \end{equation}
 \begin{equation}\label{GK-an2}
  A^\dag_{\rm GK} |n\rangle = \sqrt{e_{n+1}}
   e^{i\alpha(e_{n+1}- e_n)}|n+1\rangle,
\end{equation}
\begin{equation}\label{GK-an3}
   [A_{\rm GK}, \widetilde{A}^\dag_{\rm GK}]|n\rangle =
   (e_{n+1}- e_{n})|n\rangle,
 \end{equation}
 \begin{equation}\label{GK-an4}
   [{A}_{\rm GK}, \hat{n}] =
   {A}_{\rm GK} \qquad [{A}^\dag_{\rm GK}, \hat{n}] = -{A}^\dag_{\rm GK}.
 \end{equation}
 \subsection{{ The dual family of GKCSs as the temporally stable
    CSs of the dual of KPS CSs}}\label{sec-GK-temp}

    Now, all the necessary  tools for introducing the dual family of GKCSs
    have been prepared.
    Taking into account the results we observed about the nonlinearity nature of KPS CSs,
    comparing (\ref{kps}) with GK CSs in (\ref{GKED}),
    keeping $\hbar$ and  $\omega$ in the formulas, one may conclude that:
 \begin{equation}\label{transfer}
    e^{-i \frac{\alpha}{\hbar \omega} \HD} | z \rangle_{\rm KPS} = | z, \alpha
    \rangle, \qquad   0 \neq \alpha\in \R,  \qquad z\in \C.
 \end{equation}
   While $| z \rangle_{\rm KPS}$ states are not temporally  stable, $|z, \alpha
   \rangle$ states enjoy this property.

   Now we may outline a relatively evident physical meaning to the arbitrary
   real $\alpha$ in (\ref{GKED}) or (\ref{transfer}) as follows:
   $\alpha \equiv \omega t$,
   where by $t$ we mean the time that the operator acts on the
   KPS CSs. It should be mentioned that, in a sense this interpretation has
   been previously presented for the GKCSs in a compact form in Ref. \cite{Antoine2001}.
   But in this tutorial it is outlined in a general framework.
   If so, then $|z, \alpha \rangle$ can be considered as
   the evolution of $|z \rangle_{KPS}$.
   Therefore in a more general framework, one can  claim that the
   action of the evolution type operator
\begin{equation}\label{evolution}
  \hat{S}(\alpha) = e^{-i \frac{\alpha}{\hbar \omega} \HD},
  \qquad \hat S \hat S^\dag = \hat S^\dag \hat S = \hat{I}, \qquad 0 \neq \alpha \in
  \R,
 \end{equation}
   on any non-temporally stable CSs, makes it temporally
   stable CSs. So, $\hat{S}(\alpha)$ is a
   nice and novelty operator which falls down any generalized CS to a
   situation which it restores the temporal stability property.
   Where we stress on the fact that in (\ref {evolution}) the evolved
   Hamiltonian, $\HD$ should satisfy
   $\HD|n \rangle = \hbar \omega e_n |n \rangle$.

    At this point we are ready to find a suitable way to define the dual family
    of GKCS in a safe manner.
    First we note that the dual family of KPS CSs introduced in Eq. (\ref{kps-cs})
    has already been established in Eq. (\ref{kps-29}),
    via the following closed form  \cite{Roknizadeh2004}:
\begin{equation}\label{kps-dual}
    |\widetilde z \rangle_{\rm KPS}=\widetilde{\N}_{\rm KPS}(|z|^2)^{-1/2}
    \sum_{n=0}^{\infty}\frac{z^n}{\sqrt{
    \mu(n)}}|n\rangle, \qquad z \in \C,
\end{equation}
    where
    \begin{equation}\label{relate-dual}
    \mu(n)\equiv \widetilde{\rho}(n)=\frac{(n!)^2}{\rho(n)},
    \end{equation}
    $\mu(n)\equiv\widetilde{\rho}(n)$ is dual correspondence of
    $\rho(n)$.
    Equation (\ref {relate-dual}) expresses the relation between
    the KPS and the associated dual CSs, simply.
    Obviously $\N_{\rm KPS}$ and $\widetilde{\N}_{\rm KPS}$
    in (\ref{kps-cs}) and (\ref{kps-dual}) are the normalization
    constants may be obtained.
    Therefore, employing the formalism presented in (\ref {transfer})
    when imposed on "the dual of KPS states" in (\ref {kps-dual}),
    naturally leads one to the following
    superposition of Fock space for the {\it "dual family of GKCSs"} (we refer to as by DGKCS):
\begin{eqnarray}\label{DGKED}
   \widetilde{\hat{S}}(\alpha) |\widetilde{z} \rangle_{\rm KPS}  &=&
    e^{-i \frac{\alpha}{\hbar \omega}
    \widetilde{\hat{H}}}|\widetilde{z} \rangle_{\rm KPS}
     = \widetilde{\N}(|z|^2)^{-1/2}
    \sum_{n=0}^{\infty}\frac{z^n e^{-i \alpha  \varepsilon_n}} {\sqrt{
    \mu(n)}}|n\rangle \nonumber\\ &=& |\widetilde{z, \alpha}
    \rangle,
     \qquad z \in \C, \qquad 0 \neq \alpha \in \R,
\end{eqnarray}
    where $\widetilde{\N}=\widetilde{\N}_{\rm KPS}$
     (because of the unitarity of $\hat S(\alpha)$, which preserves the norm), is now determined as:
 \begin{equation}\label{normD}
     \widetilde{\N}(|z|^2)=\sum_{n=0}^{\infty}\frac{|z|^{2n}}{\mu(n)}.
 \end{equation}

     The special case of $\varepsilon_n=n$ will recover the canonical CSs, correctly.
     Note also that setting $\alpha=\omega t$ in (\ref{evolution})
     and (\ref{DGKED}) reduces the operators $\hat{S}(\alpha)$
      and $\widetilde{\hat{S}}(\alpha)$ to the well-known {\it time evolution
     operators} $\mathcal{U}(t)$ and
     $\widetilde{\mathcal{U}}(t)$, respectively.
     The case $\alpha = 0$ in the states in (\ref {GKED}) and (\ref {DGKED})
     will recover KPS and the corresponding dual CSs
     (which certainly are not temporally stable),
     respectively. The overlap between two states of the dual family of GKCSs
     takes the following form
\begin{equation}\label{overlap}
     \widetilde{ \langle z, \alpha }| \widetilde{z',
     \alpha^ \prime}\rangle =
      \widetilde{\N}(|z|^2)^{-1/2} \widetilde{\N}(|z'|^2)^{-1/2}
      \sum_{n=0}^{\infty} \frac{(z^* z')^n e^{-i \varepsilon_n (-\alpha
      +\alpha')}}{\mu(n)} \; ,
\end{equation}
     which means that the states are essentially nonorthogonal.

    In what follows we will observe that the states
    $\widetilde{|z, \alpha\rangle}$ introduced
    in (\ref {DGKED}) are exactly of GK type.
    It should be noticed that the produced states form a new class of
    generalized CSs, essentially other than $|z, \alpha\rangle$ in (\ref{GKED}).
    Also it is apparent that for our introduction, we have obtained directly
    the analytic representation of DGKCS of any arbitrary quantum mechanical system.
    Again using the nonlinear CSs method proposed in section 2.1,
    one can now deduce the nonlinearity function for the dual states in Eq. (\ref{DGKED}) as \cite{Roknizadeh2004}:
\begin{equation}\label{nlDGKED}
    \widetilde{f}_{\rm GK}(\alpha, \hat{n}) = e^{i\alpha(\hat{\varepsilon}_n -
    \hat{\varepsilon}_{n-1})}\sqrt{\frac{\mu(\hat{n})}
    {\hat{n}\mu(\hat{n}-1)}},\qquad  \alpha \; \rm{being \; fixed} ,
  \end{equation}
    where  the notation
    $\hat{\varepsilon}_n \equiv\mu(\hat{n})/\mu(\hat{n}-1)$ has been choosed.
    Therefore the deformed annihilation and creation operators
    of the dual system in analogue to (\ref {normalH}) may expressed explicitly as:
 \begin{equation}\label{annihil-DGKED}
     \widetilde{A}_{\rm GK}= a e^{i\alpha(\hat{\varepsilon}_n -
    \hat{\varepsilon}_{n-1})}\sqrt{\frac {\mu(\hat{n})}
    {\hat{n}\mu(\hat{n}-1)}},
 \end{equation}
 \begin{equation}\label{ceat-DGKED}
    \widetilde{A}^\dag _{\rm GK}= e^{-i\alpha(\hat{\varepsilon}_n -
    \hat{\varepsilon}_{n-1})}\sqrt{\frac {\mu(\hat{n})}
    {\hat{n}\mu(\hat{n}-1)}} a ^\dag.
 \end{equation}
   The normal-ordered Hamiltonian of dual oscillator in the same manner
   stated in (\ref{normalH}) is:
 \begin{equation}\label{Hamilt1}
   \widetilde{\HD}_{\rm GK} \equiv \widetilde{\HD}=
   \widetilde{A}^\dag _{\rm GK}\widetilde{A} _{\rm GK} =  \frac{\mu(\hat{n})}
   {\mu(\hat{n}-1)} = \frac{\hat{n}^2}{\hat{e}_n},
 \end{equation}
    which is again independent of $\alpha$.
    As a result
 \begin{equation}\label{Hamilt2}
   \widetilde{\HD} |n\rangle =
    \widetilde{\mathcal{E}}_n |n\rangle \equiv \hbar\omega\varepsilon _n |n\rangle
    =\varepsilon _n |n\rangle,
    \qquad  \varepsilon_n \equiv \widetilde{e}_n=\frac{n^2}{e_n},
 \end{equation}
   where again we use the units $\omega=1=\hbar$.
   The right equation in (\ref {Hamilt2}) illustrates clearly the relation between
   the eigenvalues between the two mutual dual systems.
   The dual family of GKCSs also are required to satisfy
   the following inequalities:
 \begin{equation}\label{enIneq}
   0 = \varepsilon_0 < \varepsilon_1 <  \varepsilon_2< \cdots <  \varepsilon_n < \varepsilon_{n+1} <
   \cdots .
 \end{equation}
    At this point a question may be arisen, to what extent can we may be sure that the
    dual family of GKCSs in (\ref {DGKED}) are of the GK type. It is easy to  investigate the four criteria and
    found the affirmative answer to this question (see for detail Ref. \cite{Roknizadeh-Tav-AIP}).
        We only imply the fact that upon the temporal stablity and  action identity
    requirements we are  lead  to the condition
  \begin{equation}\label{}
      \varepsilon_n = \frac{\mu(n)}{\mu(n-1)},
   \end{equation}
   which by conventional choice of $\mu(0) \equiv 1$, we deduce:
   \begin{equation}\label{mudual}
    \mu(n)= \varepsilon_n\varepsilon_{n-1}...\varepsilon_1
    =\Pi_{k=1}^{n}\varepsilon_k\equiv[ \varepsilon_n]!\; .
  \end{equation}
  So, the nonlinearity function can be expressed in terms of the
  eigen-values of the Hamiltonian system as follows
\begin{equation}\label{nlDGKED-e}
    \widetilde{f}_{\rm GK}(\alpha, \hat{n}) = e^{i\alpha(\hat{\varepsilon}_n -
    \hat{\varepsilon}_{n-1})}\sqrt {\frac{\hat \varepsilon_ n}{\hat n}}\; , \qquad  \alpha \; \rm{being \; fixed} .
  \end{equation}
    It should be noted that the same arguments I presented in section
    \ref{sec-Discusson} about the resolution of the identity (and the integration procedures),
    the $\alpha$ parameter (the states and the operators which depend on it),
    and the corresponding Hilbert spaces, must also be considered in
    the dual family of GKCSs built in the present section.

 \subsection{{The introduction of temporally stable or Gazeau-Klauder
    type of nonlinear CSs}}\label{sec-NL-temp}
    Let me now outline the main idea, in a general framework.
    It is believed that the property of the temporal stability is
    intrinsic to the harmonic oscillator and the systems which are
    unitarily equivalent to it \cite{Klauder2004}.
    But in what follows I shall demonstrate that how this
    important property can be restored by a redefinition of any generalized CSs
    which can be classified in the nonlinear CSs category.
    Recall that the nonlinear CSs  have been introduced
    in (\ref {nonl-cs}) do not have generally the temporal stability requirement \cite{Manko1997}.
    So, upon adding the results in the previous
    work \cite{Roknizadeh2004} and the above explanations, one may go proceed
    and introduce generally the new notion of
    {\it "temporally stable"} or {\it "Gazeau-Klauder
    type of  nonlinear CSs"} as
 \begin{equation}\label{nonl-temp-cs}
 \fl   |z, \alpha
  \rangle_{f}=\N_f(|z|^2)^{-1/2}\sum_{n=0}^{\infty}\frac{z^n e^{-i
  \alpha e_n}}{\sqrt{n!}[f(n)]!}|n\rangle,
  \qquad e_n=nf^2(n), \qquad 0 \neq \alpha \in \mathbb{R}, \quad z\in
  \mathbb{C}.
 \end{equation}
    We can also define the dual of the latter states  by the following expression:
 \begin{equation}\label{Roy-nonl-temp-cs}
 \fl    |\widetilde{z, \alpha}
     \rangle_{f}=\widetilde{\N}_f(|z|^2)^{-1/2}
     \sum_{n=0}^{\infty}\frac{z^n [f(n)]! e^{-i \alpha  \varepsilon_n}}
     {\sqrt n!} |n \rangle, \quad  \varepsilon_n
     = \frac{n}{f^2(n)}, \quad 0 \neq \alpha \in \mathbb{R}, \quad z\in
     \mathbb{C},
 \end{equation}
    which are indeed the temporally stable version of the nonlinear CSs
     introduced in \cite{royroy}.
    In both of the CSs in (\ref {nonl-temp-cs}) and (\ref
   {Roy-nonl-temp-cs}), $\alpha$ is a real constant and the normalization
   factors are independent of $\alpha$.
   Setting $\alpha = 0$ in (\ref {nonl-temp-cs}) and (\ref
   {Roy-nonl-temp-cs}), will recover the old form of Man'ko's and
   Roy's nonlinear CSs, respectively, which clearly were not temporally stable.

 \subsubsection{ Temporally stable CSs of $SU(1, 1)$ group}

   An instructive example of the families of nonlinear CSs
   is provided by the GP  and
   BG CSs, defined for the
   discrete series representations of the group $SU(1,1)$.
   Imposing the proposed formalism on BG CSs in (\ref{BG1}), then the
   {\it "temporally stable CSs of BG type associated with $SU(1, 1)$ group"}
   can be defined as:
 \begin{equation}\label{gilperCS-temp}
    |z, \alpha\rangle_{\rm GP}^{SU(1, 1)} = \N_{\rm GP}(|z|^2)^{-1/2}\sum_{n=0}^\infty
    \frac{z^n e^{-i \alpha \frac {n}{(n+2\kappa-1)}}}
    {[n!/\Gamma(n+2\kappa)]^{1/2}}|n\rangle,    \qquad |z|<1,
 \end{equation}
  where $\N_{GP}$ is a normalization factor, may be calculated.
  Analogously,
  applying the presented extension on the GP type of CSs in (\ref{gilperCS}), gives immediately
  {\it "temporally stable CSs of BG type associated with $SU(1, 1)$
  group"} as follows:
 \begin{equation}\label{bargirCS-temp}
 \fl  |z, \alpha\rangle_{\rm BG}^{SU(1, 1)} \equiv  |\widetilde{z, \alpha}\rangle_{\rm GP}^{SU(1, 1)}=
   \N_{\rm BG}(|z|^2)^{-1/2}
   \sum_{n=0}^\infty\frac{z^n e^{-i \alpha n(n+2\kappa-1)}}{[n!\Gamma(n+2\kappa
   )]^{1/2}} |n\rangle,    \qquad  z \in \C,
 \end{equation}
  where once more, $\N_{\rm BG}$ is chosen by normalization of the states.


 \subsubsection{ Temporally stable CSs of PS type and its dual}
     As it is observed, the generalized CSs
    introduced by Penson and Solomon \cite{Penson, Roknizadeh2004} in Eq.
    (\ref{ps1}) are also nonlinear with $f(n)= q^{(1-n)}$ and therefore the
  factorized Hamiltonian reads $\HD_{PS}=\hat{n}q ^{2(1- \hat{n})}$.
  It is stated in \cite{Penson} that under the
  action of $\exp(-i \hat{H}t)$ these
  states are temporally stable, where $\hat{H}=a^\dag a=\hat{n}$.
  Knowing that the latter Hamiltonian expresses only the (shifted)quantum harmonic oscillator
  with the corresponding canonical CS,
  seemingly to verify the invariance under time evolution operator,
  it may be more realistic to act the $\exp(-i \HD_{PS}t)=\exp (-i \hat{n}q ^{2(1- \hat{n})} t)$ operator on
  the states in (\ref{ps1}). Clearly by such proposition
  these states are not temporally stable.
  Moreover, the presented formalism allows one to construct the temporally stable
  CSs of PS type as follows
\begin{equation}\label{ps-temp}
  |q, z, \alpha \rangle_{\rm PS} \equiv e^{-i \frac{\alpha}{\hbar \omega} \HD_{\rm PS}}
   |q, z\rangle = \N(q, |z|^2)\sum _{n=0}^\infty \frac  {q^{\frac{n(n-1)}{2}}}{\sqrt{n!}}
    e^{-i \alpha e_n}  z^n |n \rangle,
 \end{equation}
  where $e_n=nq ^{2(1-n)}$, which the requested property  may be verified, straightforwardly.
  We have already introduced the dual of the PS states of Eq.
  (\ref{ps1}). So the temporally stable of the dual states may also be
  obtained immediately as
\begin{equation}\label{ps-temp}
   |\widetilde{q, z, \alpha} \rangle _{\rm PS}\equiv e^{-i \frac{\alpha}{\hbar \omega}
    \widetilde{\HD}_{\rm PS}}
    |\widetilde{q, z}\rangle = \widetilde{\N}(q, |z|^2)\sum _{n=0}^\infty \frac
    {q^{\frac{-n(n-1)}{2}}}{\sqrt{n!}}
     e^{-i \alpha \varepsilon_n}  z^n |n \rangle,
 \end{equation}
 where $\varepsilon_n=\frac{n}{q^{2(1-n)}}$.


    I end this subsection with some remarkable points.
 \begin{itemize}
\item
   In the light of the presented explanations
   the annihilation operator eigenstate for GKCSs and associated dual family are:
 \begin{equation}\label{aniGKED-def}
     A_{\rm GK}|z, \alpha \rangle = z |z, \alpha \rangle, \qquad
     \widetilde{A}_{\rm GK} |\widetilde{z, \alpha }\rangle = z |\widetilde{z, \alpha} \rangle.
 \end{equation}
    The deformed annihilation and creation operators $\widetilde{A}_{\rm GK}$
    and $\widetilde{A}^\dag_{\rm GK}$ of the dual oscillator algebra, satisfy the eigenvector equations:
 \begin{equation}\label{GK-da1}
   \widetilde{A}_{\rm GK}|n\rangle = \sqrt{ \varepsilon_n}
   e^{i\alpha ( \varepsilon_n-\varepsilon_{n-1})}|n-1\rangle,
 \end{equation}
 \begin{equation}\label{GK-da2}
   \widetilde{A}^\dag_{\rm GK} |n\rangle = \sqrt{\varepsilon_{n+1}}
   e^{i\alpha(\varepsilon_{n+1}- \varepsilon_n)}|n+1\rangle,
\end{equation}
\begin{equation}\label{GK-da3}
   [\widetilde{A}_{\rm GK}, \widetilde{A}^\dag_{\rm GK}]|n\rangle =
   (\varepsilon_{n+1}- \varepsilon_{n})|n\rangle,
 \end{equation}
 \begin{equation}\label{GK-da4}
   [\widetilde{A}_{\rm GK}, \hat{n}] =
   \widetilde{A}_{\rm GK} \qquad [\widetilde{A}^\dag_{\rm GK}, \hat{n}] = -\widetilde{A}^\dag_{\rm GK}.
 \end{equation}
   Upon looking on the actions defined in (\ref{GK-da1}) and (\ref{GK-da2})
   one can interpret $\widetilde{A}_{\rm GK}$ and $\widetilde{A}^\dagger_{\rm GK}$ as
   the operators which correctly annihilates and creates
   one quanta of the deformed photon, respectively.
   A closer look at the basis of the involved Hilbert space $\HH_\alpha$
   in each over-complete set $\{|z, \alpha\rangle\}$, shows that
   it spanned by the vectors
 \begin{equation}\label{spann-H}
   |n, \alpha\rangle=\frac{(\widetilde{A}^\dag_{\rm GK})^n e^{i \alpha
    \varepsilon_n }}{\sqrt {[e_n]!}} |0\rangle  \equiv |n \rangle,    \qquad
   \widetilde{A}_{\rm GK} |0\rangle =0.
 \end{equation}
   Moreover,  the $\alpha$ parameter from the basis has been omitted for simplicity.
   At last we are able to introduce the generators of the deformed oscillator
   algebra \cite{borzov95} of Gazeau-Klauder and the corresponding dual family
   as $\{A_{\rm GK}, A^\dag_{\rm GK}, \HD \}$ and $\{\widetilde{A}_{\rm GK}, \widetilde{A}^\dag_{\rm GK},
   \widetilde{\HD}\}$, respectively.
\item
   The probability distribution for
   the GKCSs is defined as:
\begin{equation}\label{distribut-Ev}
   \widetilde{\mathbf{P}}(n) = \left|\langle n |\widetilde{z,
   \alpha}\rangle\right|^2=\widetilde{\N}(|z|^2)^{-1}\frac{|z|^{2n}}{\mu(n)},
\end{equation}
   which is independent of $\alpha$ parameter.
 \end{itemize}

   Let me terminate this section with recalling that there exist
   also a set of equations such as (\ref
   {GK-da1}-\ref {GK-da4}) related to GKCSs, may be obtained just by replacing:
   $\widetilde{A}_{\rm GK}$, $\widetilde{A}^\dag_{\rm GK}$ and $ \varepsilon_n$
   with $A_{\rm GK}, A_{\rm GK}^\dag$ and $e_n$,
   respectively. The latter have been already derived by applying SUSYQM
   techniques \cite{Witten}, but re-derivation of them are very easy by our
   formalism.
   According to their results, the one-dimensional SUSYQM provides
   a mathematical tool to define ladder operators for an exactly solvable potentials \cite{Elkinani2003}.
   But the authors did not implied the {\it explicit} form of
   the ladder operators, and only the concerning actions were expressed there.
   Therefore besides the simplicity of our method, it is more complete
   in the sense that as we observed the {\it explicit} form of the
   rising and lowering operators in terms of the the standard bosonic creation
   and annihilation operators and
   the photon number (intensity of the field) have been found
   easily(see equations: (\ref {crea-anni}), (\ref{annihil-DGKED}), (\ref
   {ceat-DGKED})).


\subsection{Displacement operators associated with GKCS and
    the corresponding dual family}\label{sec-displca}
    After which  the explicit form of the
    deformed annihilation operator (and hence the annihilation operator
    definitions  for GKCSs and associated dual family according to Eqs. in
    (\ref {aniGKED-def})) have been introduced,
    now we are in the position  to extract the CSs of Klauder-Perelomov type
    for an arbitrary quantum mechanical system.
    For this purpose let introduce the auxiliary operators related to GKCSs:
 \begin{equation}\label{DisplaceGK}
   B_{\rm GK} =a \frac{1}{f_{\rm {GK}}(-\alpha, \hat{n})},
   \qquad B_{\rm GK}^\dag =  \frac{1}{f^\dag_{\rm {GK}}(-\alpha, \hat{n})} a^\dag.
 \end{equation}
   and analogously those for the dual families of GKCSs:
 \begin{equation}\label{DisplaceDGK}
   \widetilde{B}_{\rm GK} =a \frac{1}{\widetilde{f}(-\alpha, \hat{n})}, \qquad
   \widetilde{B}^\dag _{\rm GK}=\frac{1}{\widetilde{f}(-\alpha, \hat{n})} a^\dag.
 \end{equation}
   Notice that the minus sign in ($-\alpha$) of the argument of the $f$ and
   $\widetilde{f}$-functions
   is needed in both cases, since only in such cases we have
   $f^\dag_{\rm GK}(-\alpha, \hat{n})=f_{\rm GK}(\alpha, \hat{n})$.

   The constructed $f$-deformed operators in (\ref {DisplaceGK}) and (\ref{DisplaceDGK})
   are canonically conjugate of the $f$-deformed creation and annihilation
   operators  $(A_{\rm GK}, A_{\rm GK}^\dag)$ and $(\widetilde{A}_{\rm GK},
   \widetilde{A}^\dag_{\rm GK})$, respectively;
   i.e. satisfy the algebras $[A_{\rm GK}, B_{\rm GK}^\dag]=[B_{\rm GK},
   A_{\rm GK}^\dag]=\hat{I}$ and $[\widetilde{A}_{\rm GK}, \widetilde{B}^\dag_{\rm GK}]=[\widetilde{B}_{\rm GK},
   \widetilde{A}^\dag_{\rm GK}]=\hat{I}$, respectively.
   Now we have all mathematical instruments to construct the
   displacement operators for GKCSs:
\begin{equation}\label{disGKED}
   D_{\rm GK}(z, \alpha)=\exp(z B_{\rm GK}^\dag - z^* A_{\rm GK})
\end{equation}
   and for the dual family of GKCSs in a similar manner:
\begin{equation}\label{disDGKED}
  \widetilde{D}_{\rm GK}(z, \alpha) = \exp(z \widetilde{B}^\dag _{\rm GK}- z^*
  \widetilde{A}_{\rm GK}).
\end{equation}
   The actions of $D_{\rm GK}(z, \alpha)$ and $\widetilde{D}_{\rm GK}(z, \alpha)$
   in the latter equations on the vacuum state $| 0 \rangle$ yield the GKCSs and the associated dual family, up
   to some normalization constant,respectively, as we demand.
   From the group theoretical point of view,
   one can see that the sets $\{A_{\rm GK}, B _{\rm GK}^\dag, B_{\rm GK}^\dag A_{\rm GK}, \hat I\}$ and
   $\{\widetilde{A}_{\rm GK}, \widetilde{B}^\dag_{\rm GK}, \widetilde{B}^\dag_{\rm GK} \widetilde{A}_{\rm GK}, \hat I\}$
   which are respectively responsible for GKCSs and dual family of GKCSs, form the
   Lie algebra $h_4$, and the corresponding Lie group is the
   well-known Weyl-Heisenberg (WH) group. Also the action of the latter operators
   on the vacuum  are the orbits of the projective {\it non-unitary}
   representations of the W-H group \cite{AliRokTav}. It must be
   understood that as it is pointed out earlier,  neither the
   formalism in \cite{AliRokTav} nor the equivalent formalism of
   Ref. \cite{royroy} for constructing the
   dual states  have not been applied, since the states obtained from the earlier formalisms
   were not full consistent with the Gazeau-Klauder's criteria.
   Indeed  a rather new way has proposed in this tutorial, based on the central idea,
   through viewing the GKCSs $|z, \alpha\rangle$ and its dual pair
   $|\widetilde{z, \alpha}\rangle$ as
   generalization of KPS nonlinear CSs $|z\rangle_{\rm KPS}$ and its dual
   $|\widetilde{z}\rangle_{\rm KPS}$ to the two distinct temporally stable CSs, respectively.
   Speaking otherwise, the operators introduced in (\ref{disGKED}) and
   (\ref {disDGKED}) do not have the relation:
   $\widetilde{D}_{GK}(-z, \alpha)=D_{GK}(z, \alpha)=[D_{GK}(z,
   \alpha)^{-1}]^\dag$, which is the characteristics of the
   earlier formalisms.
   Also it is possible to build the following displacement type
   operator,
  $ V_{\rm GK}(z, \alpha)=\exp(z A_{\rm GK}^\dag -z^* B_{\rm GK})$
     for GKCSs, and in a similar manner,
  $ \widetilde{V}_{\rm GK}(z, \alpha)=\exp(z \widetilde{A}^\dag _{\rm GK}- z^*
  \widetilde{B}_{\rm GK})$
    for dual family of GKCSs, the actions of which on the vacuum state, yield two new sets of states.
    But it is easy to check that none of them can be classified in the Gazeau-Klauder CSs.

  By various superpositions of CSs, different nonclassical states
  of light may be constructed.
  Recently, there has been much interest in the construction as well as generation
  of these states, because of their properties in the context of
  quantum optics. Their different characteristics is due to the various
  quantum interference between summands.

  By various superpositions of CSs, different nonclassical states may be constructed.
  As an example, the even and odd CSs of canonical CSs as
  well as other classes of generalized CSs such as
  nonlinear CSs well studied in the literature \cite{Mancini}, due to
  their nonclassical features, such as squeezing, sub-Poissonian
  statistics (antibunching) and oscillatory number distribution.
  The symmetric (antisymmetric) combinations of GKCSs
  have been introduced in \cite{ElKinani2002}. Similarly using the
  introduced dual families of GKCSs, one led to the even ($+$) and odd ($-$) CSs denoted by
  $|\widetilde{z, \alpha} \rangle_\pm$ \cite{Roknizadeh-Tav-AIP}.
   The resulted states $(\pm)$ satisfy
   the eigenvalue equations$(\widetilde{A}_{GK})^2 |\widetilde{z, \alpha} \rangle_\pm =
    z^2 |\widetilde{z, \alpha }\rangle_\pm$.
   The real and imaginary Schr\"{o}dinger cat states may also
   obtained by adding and subtracting $|\widetilde{z, \alpha} \rangle$ and $|\widetilde{z^*, \alpha}
   \rangle$ ($z^*$ is the complex conjugate of $z$), respectively
   (the explicit form of them can be seen in Ref. \cite{Roknizadeh-Tav-AIP}).

 \subsection{Some physical appearances of the dual family of GKCSs}\label{sec-Examples}
   In order to illustrate the presented idea in this paper,
   let me apply the formalism on some physical examples
   which the associated GKCSs have already been known.
 \vspace{4 mm}\\
   {\it Example} $1$  {\it Harmonic oscillator:}
 \vspace{2 mm}\\
   As the simplest example one can apply the formalism to the harmonic
   oscillator Hamiltonian, whose the nonlinearity function is
   equal to $1$, hence $ \varepsilon_n = n=e_n$ which results the moments
   as $\mu(n)=n!=\rho(n)$. Note that we have considered a shifted Hamiltonian to lower the
   zero-point energy to zero ($e_0=0=\varepsilon_0$). Eventually
\begin{equation}\label{}
    |\widetilde{z, \alpha }\rangle_{\rm CCS}=e^{-|z|^2/2}
    \sum_{n=0}^{\infty}\frac{z^n e^{-i \alpha n}}{\sqrt{
    n!}}|n\rangle = |z, \alpha \rangle _{\rm CCS},
\end{equation}
    ensures the self-duality of canonical CS.
    For this example all the Gazeau-Klauder's requirements are satisfied, trivially.
 \vspace{4 mm}\\
  {\it Example} $2$  {\it P\"{o}schl-Teller potential:}
\vspace{2 mm}\\
    Interesting to this potential and its CSs is due to various application in
    many fields of physics such as atomic and molecular physics.
    The usual GKCSs for the
    P\"{o}schl-Teller potential, have been demonstrated nicely by J-P Antoine
    {\it etal} \cite{Antoine2001}. Their obtained results are as follows
 \begin{equation}\label{poshCnd}
    e_n=n(n+\nu),   \qquad  \rho(n)=\frac{n!\Gamma(n+\nu+1)}{\Gamma(\nu+1)}, \qquad \nu >    2,
 \end{equation}
   with the radius of convergence $R=\infty$.
   Consequently  substitution of the quantities of equation (\ref {poshCnd}),
   in (\ref {relate-dual}) and (\ref{Hamilt2})
   one can construct the dual of the already known CSs for P\"{o}schl-Teller potential as
 \begin{equation}\label{poshCnd-Dual2}
    | \widetilde{z, \alpha} \rangle _{\rm PT}=
    (1-|z|^2)^{(1+\nu)/2} \sum_{n=0}^\infty \frac{z^n}{\sqrt{\frac{n!
    \Gamma(\nu+1)}{\Gamma(n+\nu+1)}
    }}e^{-i \alpha  \frac{n}{n+\nu}}|
    n\rangle,
 \end{equation}
   whose radius of convergence is determined as the open unit disk.
   The overlap between two of these states when $\alpha= \alpha'$ is obtained
   from(\ref{overlap}) as:
\begin{equation}
    {}_{\rm PT}\langle \widetilde{ z, \alpha} |  \widetilde{z', \alpha }\rangle_{\rm PT}=
   \left[(1-|z|^2)(1-|z'|^2)\right]^{(1+\nu)/2}(1- z z')^{(-1-\nu)}.
\end{equation}
   To be ensure, for these dual states only investigation of the
   resolution of the identity has been done, since the other three requirements
   satisfied obviously. As required  one has to find $\varrho(x)$
   such that the moment integral
\begin{equation}\label{poshCnd-Dua3}
    \int_0^1 x^n \varrho(x) dx=\frac{n! \Gamma(\nu+1)}{\Gamma(n+\nu+1)}
\end{equation}
    holds. It may be checked that the proper weight function is determined as
    $\nu (1-x)^{\nu-1}$.

    At this point,  recall that (\ref {poshCnd}) denotes
    the eigenvalues of different Hamiltonians.
    The role of the characteristics of the dynamical system plays by the parameter
    $\nu$.
    For instance it is precisely
    the eigenvalues of the anharmonic (nonlinear) oscillator also,
    well studied in literature.
    GKCSs (and GK-CSs) have been discussed
    in Refs. \cite{ElKinani2002} and \cite{Roy2003} in details, respectively.
    In current example the parameter $\nu$ is related to
    two other parameters, namely $\lambda$ and $\kappa$ through the relation $\nu=\lambda
    +\kappa$, which determine the height and the dept of the well
    potential. While when one deals with the nonlinear oscillator it has
    another meaning; e.g. we refer to Ref. \cite{Roy2003},
    in which the interest was due to its usefulness in the study
    of laser light propagation in a {\it nonlinear Kerr medium}.
    In particular $\nu$ in the latter case is related to the
    nonlinear susceptibility of the medium. So, the obtained result in
    (\ref{poshCnd-Dual2}) can be exactly used for the
    anharmonic oscillator, too. To this end, in the next example we see
    that the case of $\nu=2$ in (\ref {poshCnd}) is the exact
    eigenvalues of the infinite potential well.
\vspace{4 mm}\\
   {\it Example} $3$  {\it Infinite well potential:}
\vspace{2 mm}\\
    The GKCSs for the infinite
    well, well established by in \cite{Antoine2001}, noting that
\begin{equation}\label{well}
   e_n=n(n+2), \qquad \rho(n)=\frac{n!(n+2)!}{2},
\end{equation}
   with the radius of convergence $R=\infty$.
   Consequently  inserting (\ref{well}) in (\ref {relate-dual}) and (\ref{Hamilt2})
   one can construct the dual of these states as
\begin{equation}\label{well-Dual2}
    | \widetilde{z, \alpha} \rangle _{\rm IW}=
    (1-|z|^2)^{3/2}     \sum_{n=0}^\infty \frac{z^n}{  \sqrt{(\frac{2}{(n+1)(n+2)}) }}
    e^{-i \alpha (\frac{n}{n+2}})  | n\rangle,
 \end{equation}
   with  radius of convergence determined as the open unit disk.
   Again the overlap between two of these states for the special case  $\alpha= \alpha'$ is
   obtained from(\ref{overlap}) as:
\begin{equation}
   {}_{\rm IW}\langle \widetilde{ z, \alpha} |  \widetilde{z', \alpha }\rangle _{\rm IW}=
   - \left[ \frac{1}{(-1+|z|^2)(-1+|z'|^2)} \right]^{-3/2}
   \frac{1}{(z z'-1)}.
\end{equation}
   To clarify the fact that these dual states are actually CSs, we only investigate the
   resolution of the identity, since  the other three requirements
   satisfied straightforwardly. For this condition we have to find
   $\varrho(x)$ such that the integral
\begin{equation}\label{}
    \int_0^1 x^n \varrho(x) dx=\frac{2}{(n+1)(n+2)}
\end{equation}
    holds. It is easy to verify that $\varrho(x)=2(1-x)$ is the
    one needs in this case.
 \vspace{4 mm}\\
    {\it Example $4\; $  Hydrogen-like spectrum:}
 \vspace{2 mm}\\
    We now choose the Hydrogen-like
    spectrum whose the corresponding CSs, has been a long-standing
    subject and discussed frequently in the literature.
    For instance in Refs. \cite{gazklau, Klauder2001} the one-dimensional
    model of such a system with the Hamiltonian $\hat H = -\omega/(\hat n +1)^2$
    and the eigen-values $E_n = -\omega/(n +1)^2$ has been considered ($\omega=m
    e^4/2$, and $n=0, 1, 2, \cdots$).
    But to be consistent with the GKCSs, as it has been done in
    \cite{Klauder2001}, the energy-levels should be shifted by a
    constant amount, such that after taking $\omega=1$ one has the
    eigen-values $e_n$ and therefore the functions $\rho(n)$ as follows
\begin{equation}\label{hyd}
    e_n = 1- \frac{1}{(n+1)^2},
    \qquad \rho(n)=\frac{(n+2)}{2(n+1)},
\end{equation}
    with unit disk centered at the origin as the region of convergence, i.e. $R = 1$.
    Therefore the related dual family of GKCSs for bound state
    portion of the Hydrogen-like atom can be constructed.
    For this purpose, take into account (\ref{hyd}) in (\ref{relate-dual})
    and (\ref{Hamilt2}), so the explicit form of
    the corresponding dual family of GKCSs for this system
    can be easily obtained as
 \begin{equation}\label{Morse-Dual2}
  \fl  | \widetilde{z, \alpha} \rangle _{\rm H} =
\left ( \frac{1}{2\sqrt{|z|^2}}\left[2I_1(2\sqrt{|z|^2}) +
\sqrt{|z|^2}
   I_2(2\sqrt{|z|^2})\right]\right)^{-1/2}
   \sum_{n=0}^\infty \frac{z^n}{\sqrt{\frac{2 n!(n+1)!}{n+2}} }
    e^{-i \alpha \left(\frac{n (n+1)^2}{n+2} \right)}  | n\rangle,
\end{equation}
     where $I_\nu(z)$ is the modified Bessel function of the first
   kind.   In this case $\tilde R =\infty$ as the radius of convergence.
   Similar to the preceding examples, we only verify the resolution of the identity.
   In the present case we have to find a
   function $\tilde{\sigma}(x)$ such that
\begin{equation}\label{hyd3}
    \int_0^\infty x^n \tilde{\sigma}(x) dx = 2\frac{
    n!(n+1)!}{n+2}.
\end{equation}
   The integral in (\ref {Meiger}) is again helpful, if we rewrite the RHS of the
   (\ref {hyd3}) as $2n![(n+1)!]^2/(n+2)!$. The suitable measure
   is then found to be
 \begin{eqnarray}\label{hyd4}
    \tilde{\sigma}(x) =  G_{0, 0}^{3, 1}
     \left( x \Big|
   \begin{array}{rr}
      0, 1, 1,& .  \\
        2   ,& .
   \end{array}
      \right).
      \end{eqnarray}
   The overlap between these states for the special case  $\alpha= \alpha'$ is
   obtained from (\ref{overlap}) in the closed form
\begin{equation}\label{overlap-hyd}
   \fl  {}_{\rm H}\widetilde{ \langle z, \alpha }| \widetilde{z',
     \alpha}\rangle _{\rm H}=
      \widetilde{\N}(|z|^2)^{-1/2} \widetilde{\N}(|z'|^2)^{-1/2}
      \frac{1}{2\sqrt{z^\ast z'}}\left(2I_1(2\sqrt{z^\ast z'})+ \sqrt{z^\ast z'}
      I_2(2\sqrt{z^\ast z'})\right),
\end{equation}
   where $\widetilde{\N}(|z|^2)=
   \frac{1}{2\sqrt{|z|^2}}\left[2I_1(2\sqrt{|z|^2}) + \sqrt{|z|^2}
   I_2(2\sqrt{|z|^2})\right]$.
\vspace{4 mm}\\
    {\it Example} $5$  {\it Morse potential:}
\vspace{2 mm}\\
    As the final physical example, I pay attention to the GKCSs for the Morse potential,
    which is the simplest anharmonic oscillator,
    useful in various problems in
    different fields of physics(for example: spectroscopy,
    diatomic and polyatomic molecule vibrations and scattering),
    can be obtained using the related quantities may be found in \cite{Popov2003}:
\begin{equation}\label{morse}
    e_n = \frac{n(M+1-n)}{M+2}, \qquad \rho(n)=\frac{\Gamma(n+1)\Gamma(M+1)}
    {(M+2)^n \Gamma(M+1-n)},
\end{equation}
    where $n=0,1,\cdots < (M+1)$.
    Therefore, taking into account (\ref{morse}) in (\ref{relate-dual}) and
    (\ref{Hamilt2}),
    the dual of these states
    can be produced as
\begin{equation}\label{Morse-Dual2}
  \fl  | \widetilde{z, \alpha} \rangle _{\rm MP}=
    \left[1+\frac{|z|^2}{2+M}\right]^{-M/2}  \sum_{n=0}^ M
    \frac{z^n}{ \sqrt{\frac{(M+2)^n
    \Gamma(n+1)\Gamma(M-n+1)}{\Gamma(M+1)} }}
    e^{-i \alpha \left(\frac{n (M+2)}{M-n+1}\right)}  | n\rangle,
\end{equation}
   where again the equation (\ref {normD}) have been used.
    Noticing that the series led to the above $\widetilde{\N}(|z|^2)$
   is now a finite series,
   makes it clear that it converges for every values of
   $x=\sqrt{|z|^2} \geq 0$, i.e. $z \in \mathbb{C}$.
   For the overlap between two of these states when $\alpha= \alpha'$
   the formula (\ref{overlap}) is not useful and one must calculate especially
   the overlap between the Morse states for themselves, because of the upper bound of the
   evolved sigma:
\begin{eqnarray}
   \fl  {}_{\rm MP} \langle \widetilde{ z, \alpha} |  \widetilde{z', \alpha
     }\rangle _{\rm MP}
  &=& \widetilde{\N}(|z|^2)^{-1/2} \widetilde{\N}(|z'|^2)^{-1/2}
     \sum_{n=0}^{M+1} \frac{(z^* z')^n} {\mu(n)}\nonumber\\
  &=&\left[\left(1+\frac{|z|^2}{2+M}\right)
     \left(1+\frac{|z'|^2}{2+M}\right)\right]^{-M/2}
     \left[\frac{2+M+zz'}{2+M} \right]^M.
\end{eqnarray}
   Verifying the resolution of the identity is necessary. As before,
   the   function $\varrho(x)$ must be found such that:
\begin{equation}\label{morse3}
    \int_0^\infty x^n \varrho(x) dx=\frac{(M+2)^n
    \Gamma(n+1)\Gamma(M-n+1)}{\Gamma(M+1)}.
\end{equation}
   Using the definition of Meijer's $G$-function and the inverse Mellin
   theorem, it follows that \cite{Mathai1973}:
 \begin{eqnarray}\label{Meiger}
    \int_0^\infty dx x^{s-1} G_{p, q}^{m, n}
     \left( \alpha x \Big|
   \begin{array}{cccccc}
      a_1, & \cdots, & a_n, & a_{n+1}, & \cdots, & a_p  \\
      b_1, & \cdots, & b_m, & b_{m+1}, &\cdots, & b_q
   \end{array}
     \right)  \nonumber\\   =   \frac{1}{\alpha ^s} \frac  {\Pi_{j=1}^{m}\Gamma
     (b_j +s) \Pi_{j=1}^{n}\Gamma (1-a_j-s)} {\Pi_{j=n+1}^{p}\Gamma (a_j+s)
     \Pi_{j=m+1}^{q}\Gamma (1-b_j-s)}.
 \end{eqnarray}
   One can find the function $\varrho(x)$ needed in (\ref {morse3}), in terms of the
   Meijer's $G$-function by the expression:
\begin{eqnarray}\label{Meiger}
 \varrho(x) = (M+2) \Gamma(M+1) G_{0, 0}^{1, 1}
     \left(  x (M+2)^{-1}\Big|
   \begin{array}{rr}
      -(M+1),& .  \\
        0   ,& .
   \end{array}
      \right).
      \end{eqnarray}
      A point is worth to mention in relation to the latter
      example, which is special in the sense that its Fock space
      is finite dimension. While in the infinite dimensional cases
      the presented formalism provide the ladder operators associated with
      each system, this is not so in the finite dimensional case.
      Because, in general it is not possible to find the algebraic
      definition for the CSs defined in the finite dimension.
      Anyway, Gazeau-Klauder definition  provides a suitable and
      simple rule to define the CSs associated to such quantum
      mechanical systems, and the presented formalism in this
      paper provides a safe way to build their dual family.
\section {\it Introducing the generalized GKCSs and the associated dual family}
    In the light of the above explanations we are now in a position to propose
    the generalized GKCSs, by which we may recover the
    GKCSs in Eq. (\ref{GKED}) and the nonlinear CSs in Eq. (\ref{nonl-cs})
    as two special cases \footnote{I must thanks to Prof. S Twareque Ali for adding this part to our works}.
    In the following scheme,
    the physical meaning of the $\alpha$ parameter
    which enters in the GKCSs will be more clear, the
    case it has  already mentioned as $\alpha=\omega t$.
\subsection {Time evolved CSs as the generalized GKCSs }
    Consider the Hamiltonian $\hat{H}$ whose eigenvectors are
    $|\phi_n\rangle$ and eigenvalues are $e_n$, such that:
\begin{equation}\label{hamlt}
   \hat{H}=\omega \sum_{n=0}^\infty e_n |\phi_n\rangle \langle
   \phi_n|, \qquad \rm{where}\quad \hat{H}|\phi_n\rangle=\omega e_n
   |\phi_n\rangle,
\end{equation}
   where $\omega$ is a constant with the dimension of energy(taking
   $\hbar=1$).
   Let $\HH$ be a separable, infinite dimensional and complex Hilbert
   space which spanned by orthonormal set $\{|\phi_n
   \rangle\}_{n=0}^\infty$.
   Also suppose $0 = e_0 < e_1 < e_2 < \cdots < e_{n} < e_{n+1} < \cdots
   $, be such that the sum $\sum_{n=0}^\infty \frac {x^n}{[e_n]!}$
   converges in some interval $0 < x \leq L$.
   For $z\in \mathbb{C}$, such that $|z|^2 < L \leq \infty$, define the generalized CSs:
\begin{equation}\label{CS}
  |z\rangle \doteq \N(|z|^2)^{-1/2}\sum_{n=0}^\infty
  \frac{z^n}{\sqrt{[e_n]!}}|\phi _n\rangle,
\end{equation}
   where $\N(|z|^2)$ being a normalization factor. As it is clear these states
   known as nonlinear CSs, with the nonlinearity function $f(n)=\sqrt{\frac{e_n}{n}}$.
   Setting $z=r e^{i\theta}$ with $r=J^\frac{1}{2}$, it is reasonable to
   write $|z\rangle \equiv |J, \theta \rangle$. Now if $d\nu$ be a
   measure which solves the moment problem
\begin{equation}\label{momp}
   \int_0^L J^n d\nu(J) = [e_n]!, \qquad \int_0^L  d\nu(J) = 1,
\end{equation}
   then these CSs satisfy the resolution of the identity
\begin{equation}\label{riden}
  \int_0^L \left[\int_0^{2\pi} |J, \theta \rangle \langle J, \theta |
  \N(J)\frac{d\theta}{2\pi} \right] d\nu(J) = I_{\HH}.
\end{equation}
   The CSs in (\ref{CS}) evolve with time in the manner
\begin{equation}\label{evolv1}
  |z, t \rangle =  e^{-i\hat{H}t}|z \rangle =
  \N(|z|^2)^{-1/2}\sum_{n=0}^\infty \frac{z^n e^{-i\omega e_n
  t}}{\sqrt{[e_n]!}}|\phi _n\rangle,
\end{equation}
    or equivalently in terms of the new variables $J$ and $\theta$
\begin{equation}\label{evolv2}
    |J, \theta, t \rangle = e^{-i\hat{H}t}|J, \theta
    \rangle =  \N(J)^{-1/2}\sum_{n=0}^\infty \frac{J^{n/2} e^{i n  \theta}e^{-i \omega e_n
    t}}{\sqrt{[e_n]!}}|\phi _n\rangle.
\end{equation}
    This larger set of GKCSs, we will call them {\it "generalized GKCSs"},
    defined for all $t$, satisfies the resolution of the identity,
\begin{eqnarray}\label{riden2}
   & &  \int_\mathbb{R}\left[ \int_0^L\left\{\int_0^{2\pi} |J,
       \theta, t \rangle \langle J, \theta, t |
       \N(J)\frac{d\theta}{2\pi}\right\} d\nu(J)\right] d
       \mu_\mathcal{B}\nonumber\\
   & =& \int_0^L\left[ \int_0^{2\pi} |J, \theta, t \rangle \langle J, \theta, t |
       \N(J)\frac{d\theta}{2\pi}\right] d\nu(J)\nonumber\\
   &=& \int_\mathbb{R} \left[ \int_0^L |J,
       \theta, t \rangle \langle J, \theta, t |
       \N(J) d\nu(J) \right] d \mu_\mathcal{B}\nonumber\\
   &=& \sum_{n=0}^\infty |\phi _n\rangle\langle \phi_n|=
       I_{\HH},
\end{eqnarray}
    where $d \mu_\mathcal{B}$ which is really a functional(not a measure)
    is referred to as the {\it  Bohr  measure},
\begin{equation}
     \langle\mu_\mathcal{B};f\rangle \doteq
     \lim_{T\rightarrow\infty}\frac{1}{2T} \int_{-T}^ T f(x)dx =
     \int_{\mathbb{R}} f(x) d\mu_\mathcal{B}(x),
\end{equation}
    and $f$ is a suitably chosen function over $\mathbb{R}$. In
    particular, if $f(x)=1$ for all $x$, then
    $\langle\mu_\mathcal{B};f\rangle=1$, so that $\mu_\mathcal{B}$
    resembles a probability measure. Therefore writing the Bohr measure as
    an integral only has a symbolic meaning.

    Setting $t=0$ in the "generalized GKCSs" of Eq. (\ref{evolv1}),
    we will recover the nonlinear CSs and if $\theta =0$ in
    the generalized CSs of Eq. (\ref{evolv2}) reduces to the GKCSs $|J, \alpha
    \rangle$ in (\ref{kps-jgama}),replacing $\gamma$ with $\alpha$,
    with $\alpha\equiv\omega t$, which the latter states satisfy the
    resolution of the identity,
\begin{equation}\label{riden3}
  \int_\mathbb{R} \left[ \int_0^L |J,
        \alpha \rangle \langle J, \alpha |
       \N(J) d\nu(J)\right] d \mu_\mathcal{B}(t) = I_{\HH}.
\end{equation}
  The generalized GKCSs $|J, \theta, t \rangle$ in (\ref{evolv2}),
  satisfy the stability condition and the action identity,
  as well as the continuity in the labels and the resolution of the
  identity,
\begin{equation}\label{}
   e^{-i\hat{H}t'}|J, \theta, t \rangle = |J, \theta, t+t' \rangle,
   \qquad \langle J, \theta, t |\hat{H}|J, \theta, t \rangle=\omega
   J,
\end{equation}
   and so do the states $|z, t\rangle$ in (\ref{evolv1}).


 \subsection {The dual family of the "generalized GKCSs"}
   Let me now write $e_n=nf^2(n)$, so using our previous results in
   the present paper,
   there are a {\it dual set} of numbers
   $\widetilde{e}_n \equiv \varepsilon_n=\frac{n}{f^2(n)}$,
   associated to the {\it dual Hamiltonian} $\widetilde{\hat{H}}$.
   Correspondingly this Hamiltonian has eigenvectors
   $|\phi_n\rangle$ and eigenvalues $\varepsilon_n$, such that:
\begin{equation}\label{hamlt}
  \widetilde{\hat{H}}=\omega \sum_{n=0}^\infty \varepsilon_n |\phi_n\rangle \langle
  \phi_n|, \qquad {\rm{where}}\quad \widetilde{\hat{H}}|\phi_n\rangle=\omega \varepsilon_n
  |\phi_n\rangle.
\end{equation}
    Also assuming that  $0 = \varepsilon_0 < \varepsilon_1 < \varepsilon_2 <
    \cdots < \varepsilon_{n} < \varepsilon_{n+1} <
    \cdots$, be such that the sum $\sum_{n=0}^\infty \frac {x^n}{[\varepsilon_n]!}$
    converges in some interval $0 < x \leq \widetilde{L}$. We can
    now define the {\it dual family} of CSs as in (\ref{CS}) by
\begin{equation}\label{CS-dual}
   |\widetilde{z}\rangle \doteq \widetilde{\N}(|z|^2)^{-1/2}\sum_{n=0}^\infty
   \frac{z^n}{\sqrt{[\varepsilon_n]!}}|\phi _n\rangle,
\end{equation}
   which are the well known(dual) nonlinear CSs of Ref. \cite{royroy}.
   The time evolution of these states are as,
\begin{equation}\label{evolv3}
  |\widetilde{z, t} \rangle =  e^{-i\hat{H}t}|\widetilde{z} \rangle =
  \widetilde{\N}(|z|^2)^{-1/2}\sum_{n=0}^\infty \frac{z^n e^{-i\omega \varepsilon_n
  t}}{\sqrt{[\varepsilon_n]!}}|\phi _n\rangle.
\end{equation}
   Again setting $z=r e^{i\theta}$ with $r=J^\frac{1}{2}$, we can
   write $|\widetilde{z}\rangle \equiv \widetilde{|J, \theta
   \rangle}$. So equivalently the states in (\ref {evolv3}) can be
   rewritten in terms of the new variables $J$ and $\theta$ as
\begin{equation}\label{evolv4}
    |\widetilde{J, \theta, t }\rangle = e^{-i\hat{H}t}|\widetilde{J,
    \theta}  \rangle =  \widetilde{\N}(J)^{-1/2}\sum_{n=0}^\infty \frac{J^{n/2}
    e^{i n  \theta}e^{-i \omega \varepsilon_n
    t}}{\sqrt{[\varepsilon_n]!}}|\phi _n\rangle.
\end{equation}
  We call this large set of states as the {\it "dual of the generalized GKCS"}.
  Setting $\theta =0$ in (\ref{evolv4}) will reduce it to the dual
  of the GKCSs we introduced in (\ref{DGKED}) with $\alpha=\omega t$.
  Provided that the moment problem
\begin{equation}\label{momp}
  \int_0^{\widetilde{L}}
   J^n d\widetilde {\nu}(J) = [\varepsilon_n]!, \qquad \int_0^{\widetilde
  {L}} d\widetilde{\nu}(J) = 1,
\end{equation}
   has a solution, we also have expressions for the resolution of
   the identity of the type (\ref{riden}), (\ref{riden2}) and
(\ref{riden3}).
   The GK criteria may immediately be verified for the dual of the generalized GKCS in Eq. (\ref{evolv4}),
   as it was down for the "generalized GKCSs" in Eq. (\ref{evolv2}).


\subsection{Generalized $f$-deformed creation and annihilation
operators}
  Define the two set of the generalized annihilation operators
\begin{equation}\label{}
  A|\phi_n\rangle = \sqrt{e_n} |\phi_{n-1}\rangle,
  \qquad \widetilde{A}|\phi_n\rangle = \sqrt{\varepsilon_n}
  |\phi_{n-1}\rangle,
\end{equation}
  and the corresponding generalized creation operators
\begin{equation}\label{}
  A^\dag|\phi_n\rangle = \sqrt{e_{n+1}} |\phi_{n+1}\rangle,
  \qquad \widetilde{A}^\dag|\phi_n\rangle = \sqrt{\varepsilon_{n+1}}
  |\phi_{n+1}\rangle,
\end{equation}
   where we recall that $\varepsilon_n\equiv\widetilde{e}_n$.
   So that the Hamiltonian and the associated dual are
\begin{equation}\label{}
  \hat{H}=\omega A^\dag A, \qquad \widetilde{\hat{H}}=\omega  \widetilde{A}^\dag
  \widetilde{A}.
\end{equation}
  For the generalized GKCSs we have
   \begin{equation}\label{GK-an1}
   A_{\rm GK}|\phi_n\rangle = \sqrt{ e_n}
   e^{i\omega t ( e_n-e_{n-1})}|\phi_{n-1}\rangle
 \end{equation}
 \begin{equation}\label{GK-an2}
   A^\dag_{\rm GK} |\phi_n\rangle = \sqrt{e_{n+1}}
   e^{i\omega t (e_{n+1}- e_n)}|\phi_{n+1}\rangle
 \end{equation}
Similarly for the dual family of generalized GKCSs the following
holds
 \begin{equation}\label{GK-an1}
     \widetilde{A}_{\rm GK}|\phi_n\rangle = \sqrt{ \varepsilon_n}
     e^{i\omega t ( \varepsilon_n-\varepsilon_{n-1})}|\phi_{n-1}\rangle
 \end{equation}
 \begin{equation}\label{GK-an2}
  \widetilde{A}^\dag_{\rm GK} |\phi_n\rangle = \sqrt{\varepsilon_{n+1}}
   e^{i\omega t (\varepsilon_{n+1}- \varepsilon_n)}|\phi_{n+1}\rangle
\end{equation}
   Then, for the states in (\ref{CS}) and (\ref{CS-dual}) we have
   \begin{equation}\label{}
   A|z\rangle =z|z\rangle, \qquad  \widetilde A|\widetilde z\rangle =z|\widetilde
   z\rangle,
\end{equation}
 and for the states in (\ref{evolv1}) and (\ref{evolv3}) we have
   clearly
\begin{equation}\label{}
   A_{\rm GK}|z, t\rangle =z|z, t\rangle, \qquad  \widetilde A_{\rm GK}|\widetilde {z, t}\rangle =z|\widetilde
   {z, t}\rangle,
\end{equation}
  Using our first formalism (nonlinear CSs method)
  for reproducing the states $|z, t\rangle$ in  (\ref{evolv1}) and $|\widetilde{z,
  t}\rangle$ in (\ref{evolv3}),  we introduce the corresponding nonlinearity
  functions explicitly as
 \begin{equation}\label{}
     f(t, n)= e^{-i\omega t
     (e_n-e_{n-1})}\sqrt{\frac{e_n}{n}},\qquad
     t \;\; \rm{  being \; fixed},
 \end{equation}
  and
  \begin{equation}\label{}
      \widetilde f(t, n)= e^{-i\omega t
      (\varepsilon_n-\varepsilon_{n-1})}\sqrt{\frac{\varepsilon_n}{n}},
      \qquad  t \;\; \rm being \; \rm fixed,
  \end{equation}
  respectively. Setting $t=0$ in the above two equations, we will obtain the nonlinearity
  functions corresponding to the states $|z\rangle$ in (\ref{CS}) and $|\widetilde z\rangle$ in
  (\ref{CS-dual}), respectively. Now the necessary tools for reproducing all the
  states (\ref{CS}), (\ref{CS-dual}), (\ref{evolv1}) and (\ref{evolv3})
  by the displacement operator definition  illustrated in section \label{sec-displca} is being in
  hand.
\section{Conclusions}
    In this tutorial I devised two different approaches deal with unifying
    the various types of generalized CSs (have been called the mathematical physics
    CSs) based
    on different analytical methods recently developed.
    In the introduction of these states the known three-fold generalizations has never been used.
    So, we outline a natural questions and try to answer it:
   is there any probable link between the
   mathematical physics generalizations of CSs in recent decade and the
   three-fold generalizations? In this relation, I found the lost
   ring which connects the above two general categories by two
   methods. a) The first is based on the
   conjecture that all of the mathematical physics generalizations in
   recent decade may be classified in the nonlinear CSs category.
   The nonlinear CSs method provides a suitable and simple way to obtain the
  (deformed) annihilation, creation, displacement and Hamiltonian
   operators, after obtaining the nonlinearity function associated
   with each set of the mathematical physics generalized CSs.
   In this manner, the connection with three-fold generalizations has
   been very clear. b) By the second approach which deals with the related
   Hilbert space of each set of generalized CSs, it is   demonstrated
   that all the so-called nonlinear CSs (which by the first approach
   have been established that it contains all the mathematical physics
  generalized CSs introduced in recent decade), as well as a large
  class of the previously known CSs in the physical literature such
  as the photon-added CSs and the binomial states will be obtained
  by change of bases in the underlying Hilbert spaces.
  Apart from the above results, using the above two approaches I found a vast new
  classes of  generalized CSs, known as the dual family related to
  each of them. Due to the fact that the two formalism can not
  produce GKCSs and the related dual family, to overcome this problem I present a rather different
  method. So, besides the above mentioned results, I introduced an
  operator, $\hat S(\alpha)$, which its action on any nonlinear CS, transfers it to a
  situation that enjoys the temporal stability property.
    We use this
  operator to obtain the GKCSs, as well as the dual family of them
  in a consistent way. Also I found a way to introduce the
  temporally stable  or Gazeau-Klauder type of nonlinear CSs.
  Finally I applied  this procedure to some quantum systems with
  known discrete spectrum, as some physical appearance and
  introduce the dual family GKCSs associated with them.
\ack{The author would like to express their utmost thanks to Dr
    R. Roknizadeh  and Prof. S. Twareque Ali
    for their intuitive comments and useful
    suggestions,  in addition to their rigorous looks at  all
    stages of works during my researches in the university of Isfahan.
    Also thanks to Dr M. Soltanolketabi, Dr. M. H. Naderi and
    all of the members of Quantum Optics Group of the
    university of Isfahan for valuable discussions.
    I would like
    to acknowledge from the university of Yazd of
    Iran for financial supports and appreciate from my wife and children for their full helps during this four
    years. Of course the presented work is indeed the the result of the collaboration of all them.}
\include{thebibliography}

\end{document}